\documentclass[aps,prd,12pt,notitlepage,nofootinbib,nobibnotes,amsmath,amssymb,tightenlines]{revtex4-1}
\usepackage{epsfig}
\usepackage{graphicx}
\usepackage{bm}
\usepackage{times}
\usepackage{braket}
\usepackage{color}
\usepackage{slashed}
\usepackage{hyperref}
\usepackage{ulem}
\usepackage{gensymb}
\usepackage{textcomp}
\definecolor{Red}{rgb}{1.,0.,0.}

\definecolor{Blue}{rgb}{0.,0.,1.}

\definecolor{Green}{rgb}{0.,1.,0.}

\definecolor{Gray}{rgb}{0.5,0.5,0.5}

\definecolor{nicered}{rgb}{0.7,0.1,0.1}
\definecolor{nicegreen}{rgb}{0.1,0.5,0.1}
\hypersetup{colorlinks,citecolor=nicegreen,linkcolor=nicered}

\begin{document}

\newcommand{\beq}{\begin{eqnarray}}
\newcommand{\eeq}{\end{eqnarray}}
\newcommand{\ben}{\begin{enumerate}}
\newcommand{\een}{\end{enumerate}}
\newcommand{\non}{\nonumber\\ }

\newcommand{\jpsi}{J/\Psi}
\newcommand{\ppa}{\phi_\pi^{\rm A}}
\newcommand{\ppp}{\phi_\pi^{\rm P}}
\newcommand{\ppt}{\phi_\pi^{\rm T}}
\newcommand{\ov}{ \overline }
\newcommand{\zerot}{ {\textbf 0_{\rm T}} }
\newcommand{\kt}{k_{\rm T} }
\newcommand{\fb}{f_{\rm B} }
\newcommand{\fk}{f_{\rm K} }
\newcommand{\rk}{r_{\rm K} }
\newcommand{\mb}{m_{\rm B} }
\newcommand{\mw}{m_{\rm W} }
\newcommand{\im}{{\rm Im} }

\newcommand{\etap}{\eta^\prime }
\newcommand{\etapp}{\eta^{(\prime)} }
\newcommand{\kks}{K^{(*)}}
\newcommand{\acp}{{\cal A}_{\rm CP}}
\newcommand{\pb}{\phi_{\rm B}}
\newcommand{\xeba}{\bar{x}_2}
\newcommand{\xsba}{\bar{x}_3}
\newcommand{\peas}{\phi^A}

\newcommand{\Dsl}{ D \hspace{-2truemm}/ }
\newcommand{\pvsl}{ p \hspace{-2.0truemm}/_{K^*} }
\newcommand{\esl}{ \epsilon \hspace{-2.1truemm}/ }
\newcommand{\psl}{ p \hspace{-2truemm}/ }
\newcommand{\ksl}{ k \hspace{-2.2truemm}/ }
\newcommand{\lsl}{ l \hspace{-2.2truemm}/ }
\newcommand{\nsl}{ n \hspace{-2.2truemm}/ }
\newcommand{\vsl}{ v \hspace{-2.2truemm}/ }
\newcommand{\xsl}{ x \hspace{-2.2truemm}/ }
\newcommand{\ysl}{ y \hspace{-2.2truemm}/ }
\newcommand{\zsl}{ z \hspace{-2.2truemm}/ }
\newcommand{\epsl}{\epsilon \hspace{-1.8truemm}/\,  }
\newcommand{\bfkk}{{\bf k} }
\newcommand{\calm}{ {\cal M} }
\newcommand{\calh}{ {\cal H} }
\newcommand{\calo}{ {\cal O} }

\def \appb{{\bf Acta. Phys. Polon. B }  }
\def \cpc{ {\bf Chin. Phys. C } }
\def \csb{ {\bf Chin. Sci. Bull.  } }
\def \ctp{ {\bf Commun. Theor. Phys. } }
\def \epjc{{\bf Eur. Phys. J. C} }
\def \ijmpcs{{\bf Int. J. Mod. Phys. Conf. Ser.} }
\def \jhep{{\bf J. High Energy Phys. } }
\def \jpg{ {\bf J. Phys. G} }
\def \mpla{{\bf Mod. Phys. Lett. A } }
\def \npb{ {\bf Nucl. Phys. B} }

\def \plb{ {\bf Phys. Lett. B} }
\def \ppn{ {\bf Phys. Part. Nucl. } }
\def \ppnp{{\bf Prog.Part. Nucl. Phys.  } }
\def \pr{  {\bf Phys. Rep.} }
\def \prc{ {\bf Phys. Rev. C }}
\def \prd{ {\bf Phys. Rev. D} }
\def \prl{ {\bf Phys. Rev. Lett.}  }
\def \ptp{ {\bf Prog. Theor. Phys. }}
\def \zpc{ {\bf Z. Phys. C}  }
\def \jpg{ {\bf J.Phys.-G-}  }
\def \ap{ {\bf Ann. of Phys}  }

\newcommand{\shc}[1]{{\color{Red} #1}}
\newcommand{\zjx}[1]{{\color{Blue} #1}}


\title{The PQCD approach towards to next-to-leading order: a short review}

\author{Shan Cheng$^{a}$ }\email{scheng@hnu.edu.cn}
\author{Zhen-jun Xiao$^b$}\email{xiaozhenjun@njnu.edu.cn}

\affiliation{$^a$ School of Physics and Electronics, Hunan University, 410082 Changsha, China, \, \non
$^b$ Department of Physics and Institute of Theoretical Physics, Nanjing Normal University, Nanjing 210023, China. \,}

\date{\today}

\begin{abstract}
In this short review we elaborate the significance of resummation in $k_T$ factorization theorem,
and summarize the recent progresses in the calculations of the next-to-leading order contributions to B meson decays
from the perturbative QCD (PQCD) approach. We also comment on the current status of the PQCD approach and
highlight some key issues to develop it in the near future for more phenomenological applications.
\end{abstract}


\maketitle


\section{Introduction}\label{Intro}

The quark confinement and the asymptotic freedom are the
two fundamental properties of quantum chromodynamics (QCD) to describe the strong interaction.
In the processes of $B$ meson decays, they are united in a convoluted formula,
where the nonperturbative and perturbative physics are written separately
in form of the universal hadron wave functions and the process-dependent hard kernels, respectively.
This idea to detach the physical amplitudes according to the acting intervals between interactions is called factorization \cite{BauerZV}.

For the two-body hadronic  $B \to M_1 M_2$ decays (here $M_i$ are generally the light mesons),  for example,
the major difficulty of the theoretical studies is still concentrated on how to calculate the hadron matrix elements
$\langle M_1M_2 \vert {\cal O}_{i} \vert B \rangle$ reliably,
where ${\cal O}_i$ are the four-quark operators in the effective Hamiltonian ${\cal H}_{eff}$ for the considered decays \cite{BuchallaVS},
\beq
{\cal H}_{eff} &=& \frac{G_F}{\sqrt{2}} \Big\{
\sum_{i=1}^{2} C_i(\mu) \big[ V_{ub}V_{us}^\ast {\cal O}_i^u(\mu) + V_{cb}V_{cs}^\ast {\cal O}_i^c(\mu) \big]  \non
&-& V_{tb}V_{ts}^\ast \sum_{j=3}^{10} C_j(\mu) {\cal O}_j(\mu) -
V_{tb}V_{ts}^\ast \big[ C_{7\gamma}(\mu) {\cal O}_{7\gamma}(\mu) + C_{8g}(\mu) {\cal O}_{8g}(\mu)\big] \Big\} \,.
\label{eq:hamiltonian}
\eeq
With the concept of the  "generalized factorization (GF) approach" \cite{BauerBM},
a phenomenological way is to introduce a small set of parameters to parameterize the non-factorizable effects \cite{AliEB,AliGB,ChenNXA}.
Based on some kind of flavour symmetries (like isospion, $SU(3)_f$ etc.,) \cite{SavageUB,ChauAY}
and Wick contraction \cite{BurasRA}, another phenomenological approach were proposed
to fit the existing data with a set of invariant subamplitudes which are further used to predict more channels.
The QCD based approaches to handle such kinds of problems are usually discussed
in the heavy quark limit and implemented by the heavy quark expansion,
such as the light-cone sum rule (LCSR) \cite{KhodjamirianMI,KhodjamirianPK,KhodjamirianEQ,KhodjamirianWN},
the QCD factorization (QCDF) \cite{BenekeBR,Beneke2000RY,BenekeRY,BenekeWA}
and the soft-collinear effective theory (SCET) \cite{Bauer:2000yr,Bauer:2001cu,Bauer:2002aj,Chay:2003ju},
and the perturbative QCD (PQCD) factorization approach \cite{KeumPH,KeumWI,LuEM,KeumMS}.
For the recent progresses in QCDF/SCET, we suggest the reader to see Refs. \cite{BecherODA,BenekeWFA} for details.
\begin{table}[thb]
\vspace{-0.5cm}
\caption{A diagrammic summary of different QCD-based approaches to study $B \to \pi$ form factor.}
\vspace{-0.5cm}
\begin{center}
\begin{tabular}{|c||c|c|c|c|} \hline
Approach & LCSRs (B meson DAs) & QCDF/SCET & PQCD \\
\hline\hline
Formula & $\Pi_i^{OPE}(Q^2, s_0^B, M^2)/(f_B m_B)$ & $C_i^{A0}(Q^2) \, \xi_{B\pi}(Q^2) + C^{B1}_i(Q^2,\omega) $ &
$\phi_B(\omega) \otimes H_i(Q^2, \omega, \nu) \otimes \phi_\pi(\nu) $ \\
& & $ \cdot \phi_B(\omega) \otimes J_i(Q^2,\omega, \nu) \otimes \phi_\pi(\nu)$ & \\
\hline
    \begin{minipage}{0.08\textwidth}
    \begin{centering}
Diagrams (LO)
    \end{centering}
    \end{minipage} &
    \begin{minipage}{0.25\textwidth}
    \begin{centering}
      \includegraphics[height=5cm,width=3.5cm,angle=0]{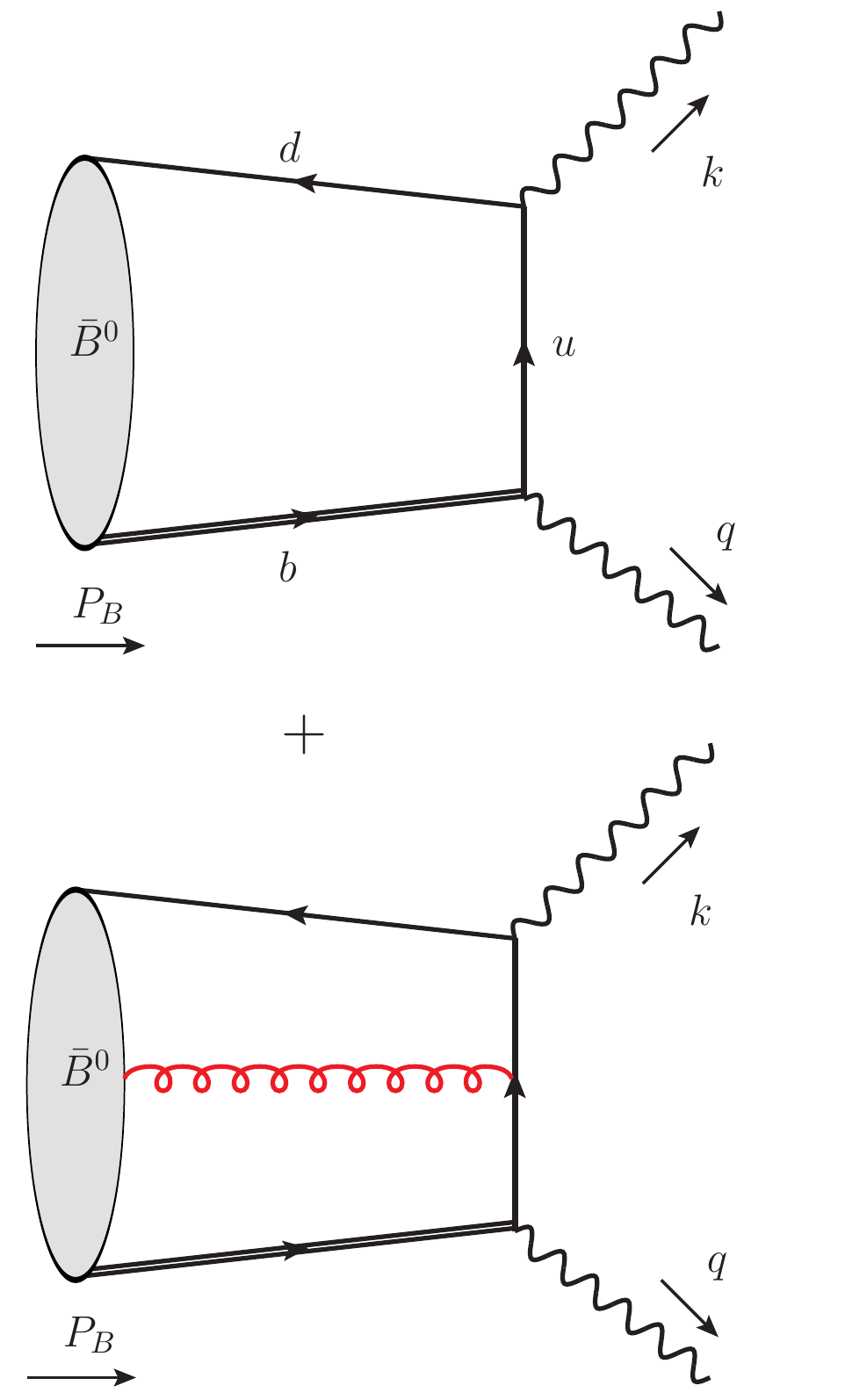}\\
      \end{centering}
    \end{minipage}  &
    \begin{minipage}{0.25\textwidth}
    \begin{centering}
      \includegraphics[height=5cm,width=3.5cm,angle=0]{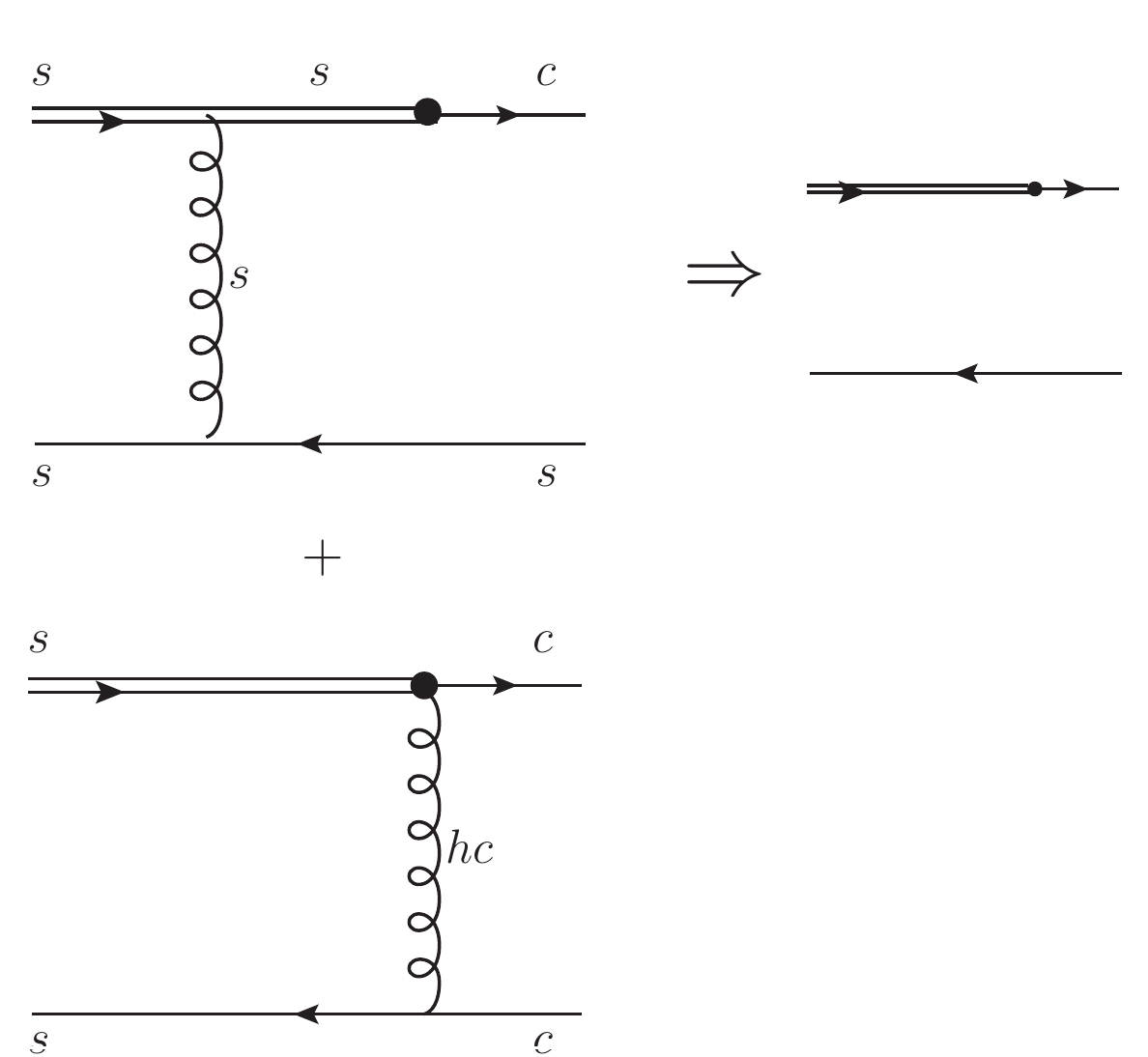}\\
     \end{centering}
     \end{minipage}  &
    \begin{minipage}{0.25\textwidth}
    \begin{centering}
      \includegraphics[height=5cm,width=3.5cm,angle=0]{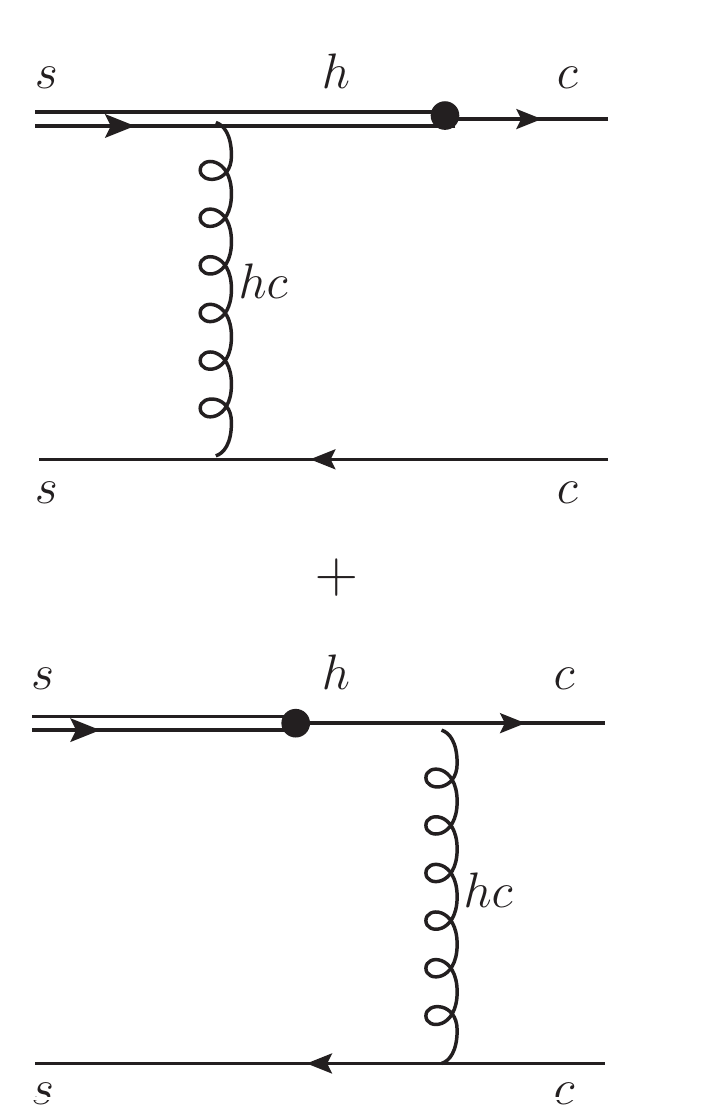}\\
    \end{centering}
    \end{minipage}  \\
    & \cite{Khodjamirian:2006st,Gubernari:2018wyi} &     \cite{Beneke2000RY,BenekeRY} &     \cite{KurimotoZJ,LuNY} \\
&&&\\
    \begin{minipage}{0.08\textwidth}
    \begin{centering}
Diagrams (NLO)
    \end{centering}
    \end{minipage} &
    \begin{minipage}{0.25\textwidth}
    \begin{centering}
      \includegraphics[height=5cm,width=3.5cm,angle=0]{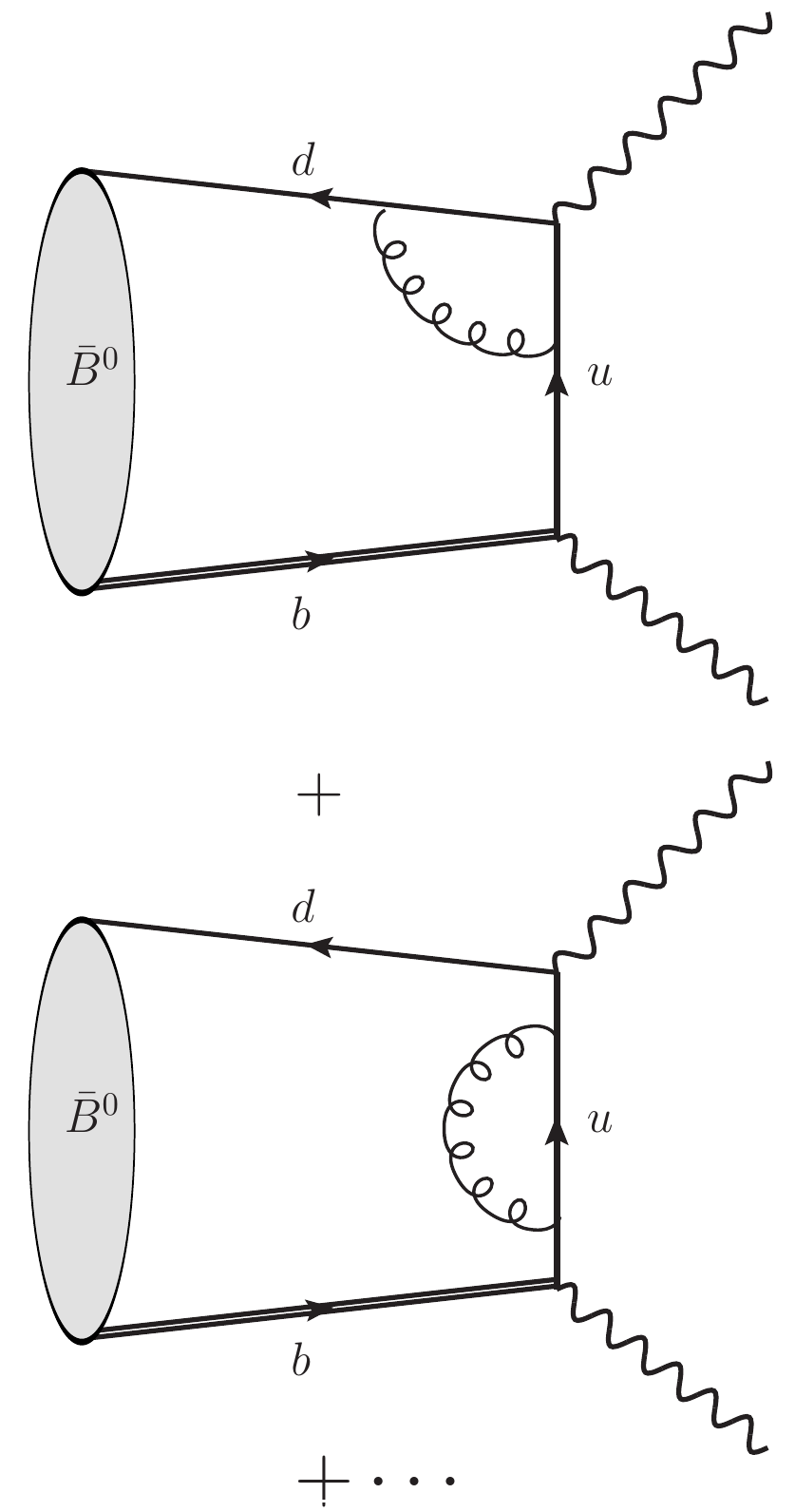}\\
       \end{centering}
    \end{minipage}  &
    \begin{minipage}{0.25\textwidth}
    \begin{centering}
      \includegraphics[height=5.5cm,width=4cm,angle=0]{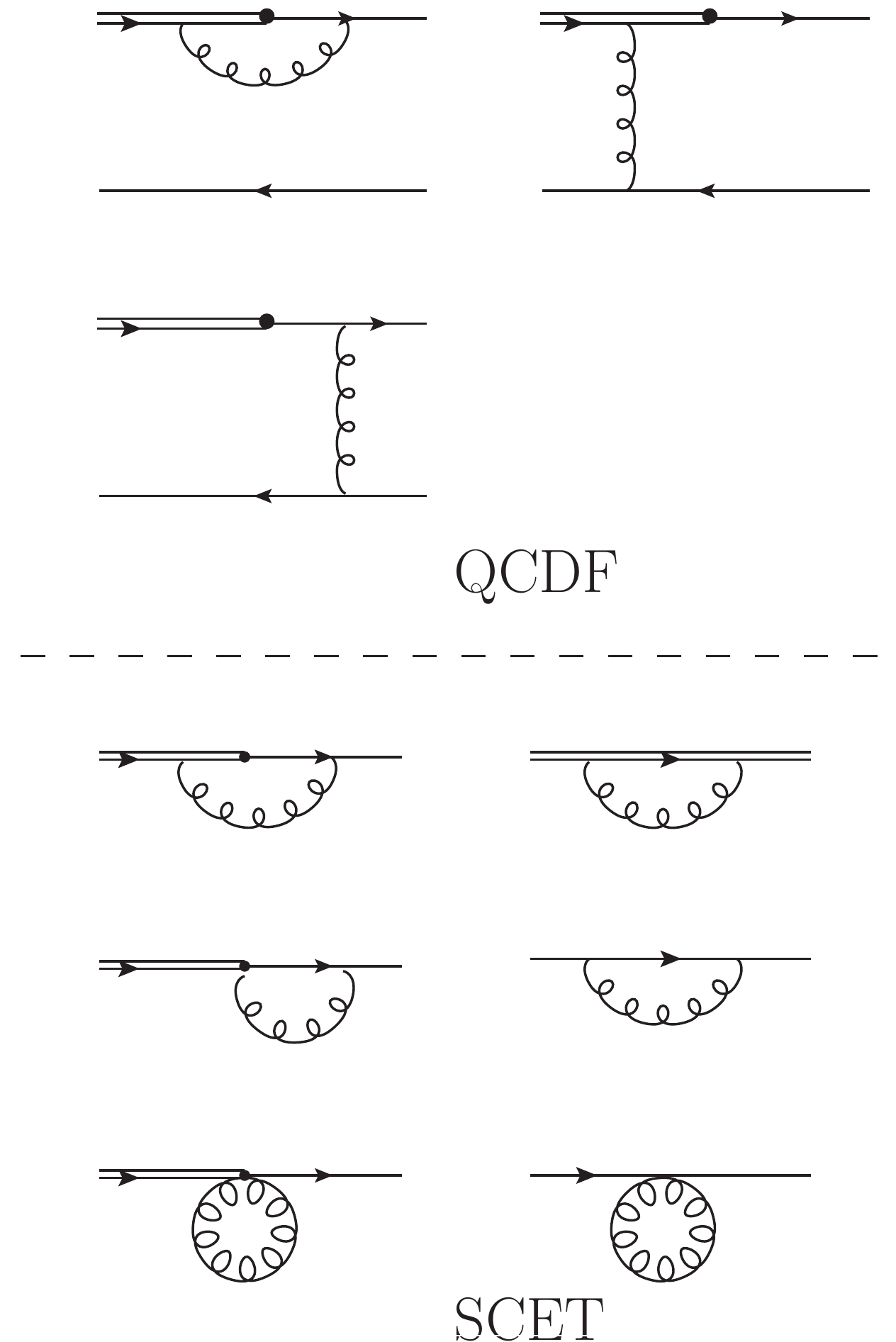}\\
     \end{centering}
     \end{minipage}  &
    \begin{minipage}{0.25\textwidth}
    \begin{centering}
      \includegraphics[height=5.5cm,width=4cm,angle=0]{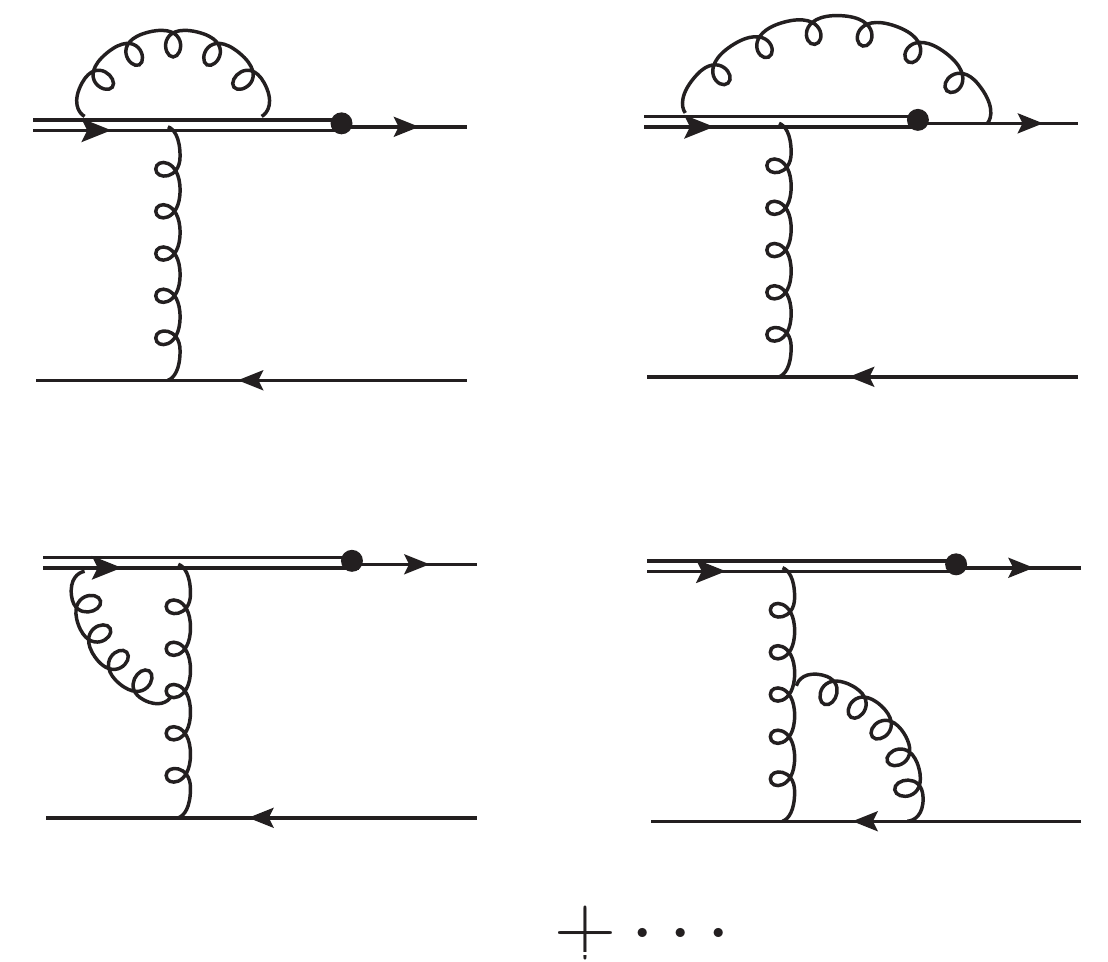}\\
    \end{centering}
    \vspace{6mm}
    \end{minipage}  \\
    & \cite{Wang:2015vgv,LuCFC} &     \cite{BenekeWA,BenekeRC} &     \cite{LiNK,ChengFWA} \\
&&&\\
    \begin{minipage}{0.08\textwidth}
    \begin{centering}
Diagrams (NNLO)
    \end{centering}
    \end{minipage} &
    \begin{minipage}{0.25\textwidth}
    \end{minipage}  &
    \begin{minipage}{0.25\textwidth}
    \begin{centering}
      \includegraphics[height=4cm,width=4.5cm,angle=0]{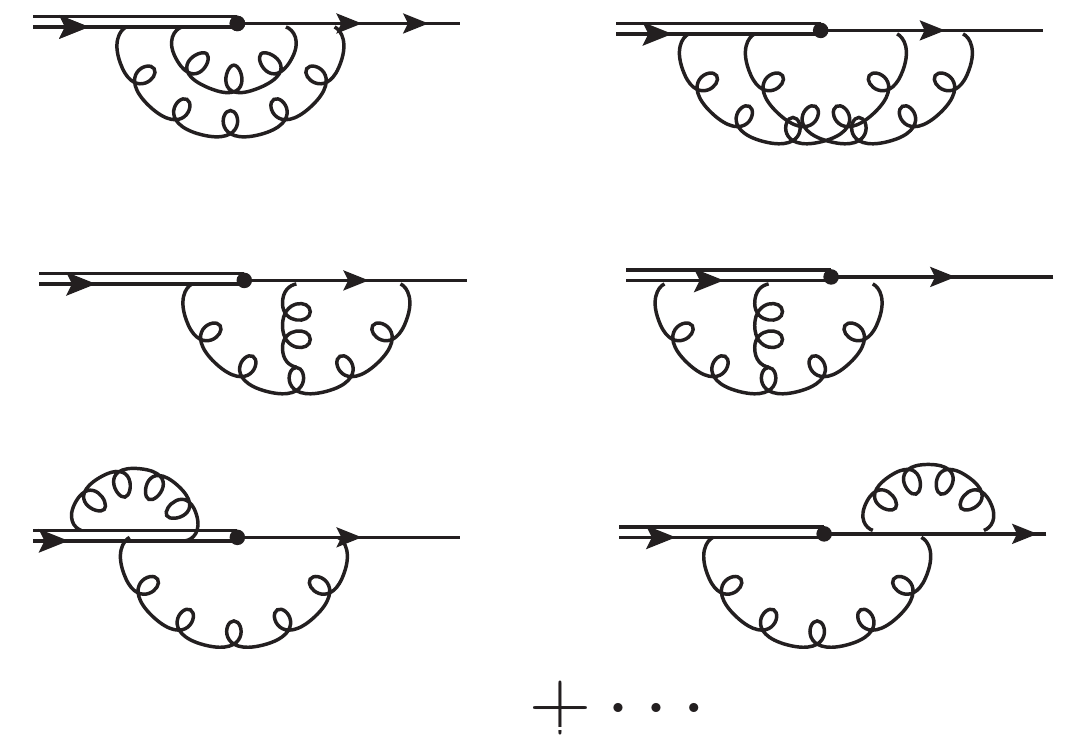}\\
     \end{centering}
     \vspace{2mm}
     \end{minipage}  &
    \begin{minipage}{0.25\textwidth}
    \end{minipage}  \\
    & &   \cite{Beneke:2009ek,Bell:2015koa,Bell:2019qya,Bell:2020qus} & \\ \hline
\end{tabular}
\end{center}
\vspace{-6mm}
\label{tab1}
\end{table}
We here highlight only the differences of their basic physical statements by taking the $B \to \pi$ transition as an  example,
\begin{itemize}
\item[(a)] The LCSR starts from the correlation functions and it believes that the heavy-to-light form factors are manipulated by the soft dynamics,
which indicates the asymmetry configuration of quark and antiquark in mesons concerning the QCD processes with large momentum transfer.
\item[(b)] In the QCDF/SCET approach, the soft-hard dynamics consider the end-point divergence into an universal soft factor $\zeta_{B \pi}$
associated with the hard matching coefficients $C_i^{A0}$, simultaneously,
the perturbative corrections is the convolution of light cone distribution amplitudes (LCDAs)
and hard function $T = C_i^{B1} \times J_i$, with $C_i^{B1}$ and $J_i$ being the hard and hard-collinear matching coefficients, respectively.
\item[(c)] With the conception of resummations, the PQCD approach states that the $B \to \pi$ form factor is a perturbative physical quantum
which is dominated by the hard dynamics, the end-point divergence is regularized by the transversal momentum of external quark lines,
and the large logarithms are resummed to a Sudakov factor in an exponent which highly suppresses the soft contribution.
\end{itemize}
In table \ref{tab1}, taking again the $B \to \pi$ transition form factor for example,
we show a diagrammic summary of the development of these three approaches to high orders of QCD coupling.
Besides the approaches mentioned above, we are pleasure to notice that
the lattice QCD (LQCD) have also been developed to simulate the heavy-to-light transition form factors
\cite{Dalgic:2006dt,Horgan:2013hoa,Flynn:2015mha,Lattice:2015tia,Lattice:2015rga},
and we suggest the review paper \cite{Kaneko:2018rng,Lattice-19}.

In this short review, we concentrate on the PQCD approach and its recent progresses towards to the next-to-leading order (NLO) accuracy.
This paper is arranged as follows.  Section \ref{PQCD} introduces briefly the basic ideas of the PQCD approach.
In section \ref{Progress}, we review the recent progresses of the PQCD approach
towards to the NLO accuracy from the views of the twist expansion, QCD radiative correction and the resummation effects in meson wave functions,
we also sum up the implementation of NLO corrections to exclusive $B$ decays.
In section \ref{Future} we discuss the key issues to improve the PQCD approach.
A summary will be given in the last section.

\section{Outline of the PQCD approach}\label{PQCD}

In this section, stemming with the understanding of large soft logarithms by the technique of the resummation,
we briefly introduce the basic ideas of the PQCD approach to treat the pion electromagnetic (e.m.) form factors and the exclusive $B$ decays.

\subsection{Resummation}

The Sudakov resummation \cite{SudakovSW} provides a well known possible solution \cite{Collins}
to do the high-order calculation and to convince the all-order argument of exclusive QCD processes,
by realising the fact that the elastic scattering amplitude of one or more isolate colored quarks, in terms of large soft logarithms,
is suppressed by means of an exponent after resuming to all order/considering all possible soft radiative corrections \cite{LepageFJ,LandshoffPB}.
The understanding of Sudakov effects took three important steps in sequence.

\begin{description}
\item[$(1)$ The origins of singularity logarithms]
For the sake of simplicity, here we introduce the singularity logarithms in the axial gauge.
The leading (double) logarithms come from the quark self-energy corrections when the transversal momentum
of exchanged gluon being in $q_{T} \leqslant 1/b$,
here $b$ is the conjugate transversal separation between the two independent scatterings with $1/b$ acting as an infrared cutoff.
Resummation in the form of Sudakov factor suppresses the soft behaviour to be close to,
but still somewhat larger than, the single hard scattering contribution \cite{MuellerSG,SenSD}.
In the same infrared region of transversal momentum,
the sub-leading (single) logarithms originate from both the self-energy and the vertex corrections of the external quark lines.
Their evolutions are derived straightly by the renormalization-group (RG) equation and
the behaviours tendency to force $b$ to be zero in the factorizable scattering amplitude.

\item[$(2)$ The resummation for elastic quark scattering]
All the logarithms in quark-quark elastic scattering can be resumed by the developed RG approach \cite{SenBT,CollinsUK,CollinsWA},
showing that the elastic scattering of an isolate colored parton is suppressed at high energies by radiative corrections.
When the radiative gluon with momentum $q$ is also on-shell as well as the constituent quark pairs,
the collinear corrections are power suppressed,
and the "soft" integral in the vicinity of $q^\mu = 0$ with $q_T \ll k_{iT}$ gives the leading singularity correction to the hard-scattering,
here $k_{iT}$ is the transversal momentum of quarks.
We can then apply the Ward identity to decouple the soft lines $S$ ($q^\mu = 0$) from the jet subdiagrams $J_i$ ($q^\mu \| k_i$),
with intermediating their connection by the eikonal lines along the jet direction.

\item[$(3)$ The resummation for elastic hadrons scattering]
In Ref. \cite{BottsKF},  Botts and Sterman resummed both the leading and nonleading logarithmic corrections to the elastic meson-meson scattering,
the resulting Sudakov factor decreases the inverse power of the momentum transfer in the divergence amplitude \cite{MuellerSG}.
The factorization theorem of elastic meson-meson scattering is proposed at leading logarithm approximation
with introducing the transversal momentum dependent (TMD) quark wave function.
Based on the factorization, the asymptotic behaviour at high energy and fixed angle is calculable
and a simple saddle point approximation on the $b$ integral is suggested for the phenomenological applications.
\end{description}

\subsection{Pion electromagnetic form factor}

At leading order (LO) with leading twist pion distribution amplitude (DA),
due to the symmetric momentum distribution of inner valence quarks,
there is no end-point singularity to pion e.m. form factor\cite{IsgurIW,HuangRD}.
The term proportional to the small transversal momentum $\mathbf{k}_T$ can be regarded as a power suppressed correction
at fixed momentum fractions,
\beq
1/(x_1x_2Q^2+\mathbf{k}^2_T) \sim 1/(x_1x_2Q^2) - \mathbf{k}^2_T/(x_1x_2 Q^2)^2 \,.
\label{eq:pi-ff-lo}
\eeq
The power suppressed (second) term in the above expression is usually neglected,
and sometimes it is taken into account by introducing a simple infrared cut.
But in fact, this term is more singular in the $x$ integral and the leading role of $k_T$ is highlighted at the endpoints regions.
For the pion e.m. form factor, the singularity begins to appear at subleading twist and/or the  NLO level of the strong coupling.

With retrieving the transversal momentum that flows from the pion wave functions through the hard scattering amplitude,
the factorization is reformulated under the PQCD approach \cite{LiNU}
in a convolution of the TMD wave functions and the hard scattering amplitudes,
{\footnotesize
\beq
F_\pi(Q^2) = \int_0^1 dx_1 \, dx_2 \int \frac{d^2\mathbf{b}_{1}}{(2\pi)^2} \frac{d^2\mathbf{b}_{2}}{(2\pi)^2} \,
\phi(x_2, \mathbf{b}_{2}, p_2, \mu) \otimes H (x_1, x_2, \mathbf{b}_{1}, \mathbf{b}_{2}, Q, \mu) \otimes \phi(x_1, \mathbf{b}_{1}, p_1, \mu) \,.
\ \ \label{eq:pi-ff-PQCD}
\eeq }
The factorization scale is chosen at the largest virtuality in the hard scattering $\mu = \textrm{Max}(\sqrt{x_1Q^2}$, $ 1/b_1, 1/b_2)$.
The pion wave functions absorb all large logarithmic enhancements at large $b_i$ extended between quarks via the exchanged soft gluon,
in which the divergent radiative corrections $\ln^2(Qb)$ and $\ln(Qb)$ in the endpoint regions are resummed to the $k_T$ Sudakov exponent
with exhibiting the suppression at large $b_i$.
After integrating over the transversal momentum with $k_T$ Sudakov factor,
the endpoint regions of longitudinal momentum fractions generate the second type of large logarithms $\ln^2[(1-x)Q]$ and $\ln[(1-x)Q]$
for the parton DA, the threshold resummation shows an enhancement at large $N$ limit,
equally saying, an enhancement at small $b_i$ with considering the $k_T$ resummation together \cite{LiIS}.
Besides the end-point enhancement, radiative corrections at higher orders may also generate the third (joint) logarithms
$\ln(x_1x_2Q^2b^2)$ in the hard kernel when $b$ is large while $x_1x_2Q^2$ is small \cite{LiIS,FengQV},
which is partly suppressed by the Sudakov exponent in pion wave functions.
With the $k_T$ resummation, the self-consistency of perturbative theorem is repaired between the strong coupling
and the large logarithms (e.g., between $\alpha_s(\mu)$ and $\ln (x_1x_2 Q^2/\mu^2)$).
The numerics shows that the PQCD begins to be self-consistent to describe the pion e.m. form factor when $Q^2 \gtrsim 4$ GeV$^2$.

\subsection{Semileptonic B decays}

With employing the heavy quark effective field theory (HQET) \cite{GeorgiUM,FalkYZ},
the Efremov-Radyushkin-Brodsky-Lepage theory of hard elastic scattering \cite{EfremovQK,LepageZA}
and the exponentiation of Sudakov double logarithms are integrated together
to calculate the semileptonic $B\to \pi l \bar{\nu}$ decay \cite{AkhouryUW}.
The soft divergences originated from quark-gluon ($\overline{b}g$) interactions are subtracted into the
standard definition of $B$ meson wave function in terms of the path-ordered eikonal phase,
then infrared (IR) safety is obtained for the LO decay amplitude.
At NLO, the radiative correction from the exchanged gluon to the eikonal phase generates soft divergence and induce the double logarithm.
The resummation effect forces the virtual gluon to be off-shell by an amount of $\sqrt{2} \Lambda_{QCD} m_b$,
yielding to an acceptable perturbative strong coupling of the leading correction.
In this standard formalism of perturbative QCD to calculate the hard scattering subdiagrams, all the particles are supposed to be far off-shell,
and the subtraction of an on-shell virtual heavy quark propagator from the hard scattering amplitudes
results in the much smaller prediction (by one power) contrasting to the data.

To resolve this deviation, a modified approach, denoted by PQCD, is developed
by picking up the transversal momentum of valence partons in both initial and final state hadrons.
The off-shellnesses of partons are regularized by $k_{T}^2$ and the eikonal IR subtraction is not needed again for the heavy quark field.
In the PQCD version, the large correction is manifested by $\ln (k_T/m_B)$,
the overlap region of soft and collinear momentum gives double logarithms to be resummed into the Sudakov factor.
The complete Sudakov exponent decreases quickly in the large $b_i$ region and vanishes as $b_i  > 1/\Lambda_{QCD}$
($b_i$ is the conjugate extent of transversal momentum),
forcing the hard gluon exchanged between valance quarks to be off shell by $\sim 8 \Lambda_{QCD} m_b$,
much larger than the value $\sqrt{2} \Lambda_{QCD} m_b$ as estimated in previous works \cite{IsgurIW,HuangRD},
and ensuring the hard scattering mechanism regardless of what's the values of $x_i$ are.
The differential decay rate of $B \to \pi l \bar{\nu}$ predicted in PQCD \cite{LiCKA,LiIU,SL-14} is comparable to the experimental data.

\subsection{Three-scale factorization in two-body hadronic $B$ decays}

Inspired by the measurements at CLEO for the exclusive charmed $B$ meson decays \cite{AlamBI},
the Bauer-Stech-Wirbel (BSW) \cite{BauerBM} approach based on the GF hypothesis
is implemented to extract the Wilson coefficient parameters $a_1$ and $a_2$ \cite{ChengFR},
while the results show an intension to describe the measurements.
In Ref.~\cite{LiKN1}, the PQCD approach is applied in $B \to \pi\pi$ decays with showing a reliable prediction.
To derive the scale independent amplitudes of two-body nonleptonic $B$ decays in the PQCD,
a three-scale factorization theorem is instructed \cite{ChangDW}.
The ultraviolet (UV) logarithms $\ln (M_W/t)$ and $\ln (t/\Lambda_{QCD})$ are summed by RG method to give the two-stage evolutions,
in which the physics stretched from scale $M_W$ down to $t$ is collected in the Wilson coefficients $C_i(M_W,t)$ and
the physics from scale $t$ down to $\Lambda_{QCD}$ is included in the hadronic matrix element $ H(t, \mu) \phi(b,\mu)$.
Supplementing with the Sudakov resummation of the IR double logarithm,
the three-scale factorization formula is quoted as
\beq
H_{r}(M_W, t ) \otimes H(t,\mu) \otimes \phi(x,P^+,b,\mu) &=& \sum_{i} C_i(M_W,t) \otimes H_i(t,t) \otimes \phi(x,b,1/b)\non
&& \hspace{-1cm}\cdot  \mathrm{Exp}\Big[ -s(P^+,b) - \int_{1/b}^{t} \frac{d\overline{\mu}}{\overline{\mu}} \gamma_\phi(\alpha_s(\overline{\mu}))\Big] \,,
\label{eq:PQCD}
\eeq
where $e^{-s}$ and $\gamma_\phi$ are the Sudakov factor and the anomalous dimension of  the wave function, respectively.
The PQCD is then extended to explain the $B \to D^{(\ast)} \pi(\rho)$ data \cite{YehRQ, LiUN},
where the four types of typical subdiagrams (say, factorizable and nonfactorizable emission, factorizable and nonfactorizable annihilation diagrams)
and the soft gluon corrections are first studied systematically.
The gauge invariance and IR finiteness of the PQCD approach to deal with the exclusive nonleptonic $B$ meson decays \cite{ChengGS}
are proofed explicitly at tree level\footnote{With considering the QCD corrections,
the additional IR subtraction terms generated by the renormalization mixing between the evanescent operators and physical operators
should be taken into account, which make the proof of gauge invariance and IR finiteness extremely challenge \cite{Wang:2017ijn,Gao:2019lta}.}. 

The PQCD approach embodies the advanced techniques of resummation and factorization,
which capture the small transversal extent of the hadrons in a hard QCD process and
absorb the residual IR divergence into the definition of wave functions, respectively,
and make sure of the application of perturbative calculation.

\section{Recent progresses towards to the NLO accuracy }\label{Progress}

\begin{figure}[tb]
\vspace{-5cm}
\centerline{\epsfxsize=18cm \epsffile{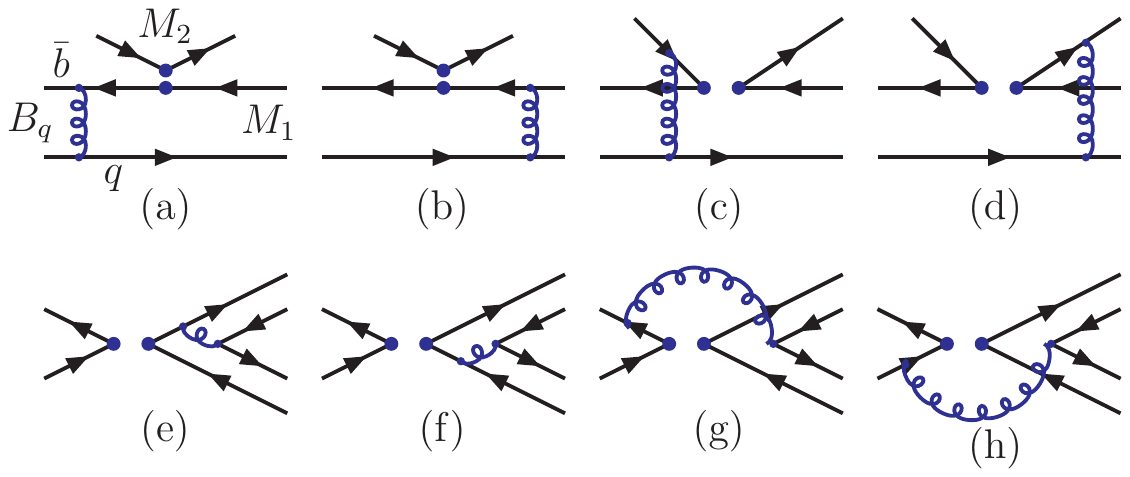}}
\vspace{-16cm}
\caption{The typical LO Feynman diagrams for the charmless two-body hadronic $B_q\to M_1 M_2$ decays in the PQCD approach.}
\label{fig:fig1}
\end{figure}

In the PQCD approach,   the typical LO Feynman diagrams for $B_q \to M_1 M_2$ decays with the spectator quark $q=(u,d,s,c)$ are illustrated
in Fig.~\ref{fig:fig1}:   the factorizable  emission  diagrams  (a,b),
the non-factorizable spectator  (c,d), and the annihilation ones (e-h).
During the past twenty years,  the PQCD factorization approach at the full leading order are wildly used  to calculate the various kinds of  $B/B_s/B_c$ meson decays,
and  has played  an important role in understanding the  relevant experimental measurements as reported by BaBar, Belle and
LHCb Collaborations \cite{Li2003,Book2013,Bfactory,HFLAV17,HFLAV19,pdg2020}.
But we know that the LO PQCD approach is just the beginning and we have to include the high order corrections in the perturbative calculations order by order,
no matter from the considerations of the precision measurements or form the side of theoretical self-consistency.

\subsection{Factorization and Form factors}

The IR safety of full amplitudes is one of the prerequisites for the perturbative calculation.
There are basically two types of IR divergences, say, the soft and collinear divergences
emerging when the radiative gluon momentum have the behaviors $l^\mu \sim (\Lambda, \Lambda, \mathbf{\Lambda_T})$
and $l^\mu \sim (Q, \Lambda^2/Q, \mathbf{\Lambda_T})$, respectively.
For convenience the light cone variables are adopted and $\Lambda$ is a small hadron scale.
The IR divergence is usually judged in an intuitive way:
the soft divergence would be generated when a gluon exchanges between two on-shell quark lines,
and the collinear divergence is usually associated with a gluon emission from a massless quark with momentum $p = (Q, 0, \mathbf{0_T})$.

Factorization of the QCD IR divergence is performed in sequence of momentum, spin and color spaces.
The final result is written in a separation of the fermion and color currents,
in which all the residual IR divergences are absorbed into hadron wave functions with remaining the hard amplitudes IR safety.
Saying more technical details, the eikonal approximation is utilized in the leading IR regions
to detach the soft and collinear gluons exchanged between the particle lines in the hard amplitude and in the wave functions,
the Fierz identity is inserted in the full amplitude to spread out the fermion current into different twists,
and finally the Ward identity is indispensable to sum over all possible color structures
and to prove the factorization theorem in a non-Abelian gauge formula.

In Ref.~\cite{LiHH}, the collinear factorization is derived at leading twist for the elastic scattering processes
of pion transition and e.m. form factors under the convariant gauge.
The proof is then extended to the two-parton twist three case \cite{NagashimaIW}.
It is shown that the soft divergences arose from different subdiagrams cancel with each other,
and the sum of collinear divergences at quark level is equal to the same type divergence
in the universal gauge invariant definition of pion meson wave function,
e.g., for the two-particle matrix element with spinor operator $\gamma_5\gamma_\mu$,
\beq
\langle 0 \vert \bar{u}(z) (\gamma_5 \gamma_\mu) \, \mathrm{P \, exp}\big[ -ig_s \int_0^z ds \cdot A(s) \big] d(0) \vert \pi^-(p) \rangle
= i p_\mu f_\pi \int_0^1 \, dx \, e^{-i \bar{x} p z}
\big\{ \phi_\pi(x) + \cdots \big\} \,,
\label{eq:pi-wf}
\eeq
where $\phi_\pi$ is the leading twist light-cone distribution amplitude (LCDA) and the ellipsis denotes the high twist (four and above) LCDAs.
The proof of factorization is firstly obtained at NLO and then presented to all orders by induction with employing the Ward identity.
Under the heavy quark limit, the factorizations of semileptonic radiative and transition $B$ decays are also derived \cite{LiHH}.
The color cloud enveloped over $b$ quark from the light spectator quark indicates that only the soft type of IR divergence appears,
which is absorbed into the gauge-invariant $B$ meson light-cone wave functions.

The $k_T$ factorization theorem of exclusive processes was rigorously proved in Ref.~\cite{NagashimaIA} at leading power,
by assuming that the on-shell valence partons carry only longitudinal momenta initially,
and acquire the small transversal component of momenta when a collinear gluon exchange happens.
The full amplitude is arranged in a convolution of the hard kernel with the LCDAs in $b$ parameter space.
Taking the pion e.m. scattering $\pi \gamma^\ast \to \pi$ for an example,
\beq
G^{(N+1)}(x,b) =  \sum_{n,m} \sum_{i=0}^{N+1} \sum_{j=0}^{N+1-i} \phi^{(i)}_n(x,\zeta,b) \otimes H^{(N+1-i-j)}(\zeta,b) \otimes \phi^{(j)}_m(x,\zeta,b) \,,
\label{eq:NLO-KH}
\eeq
where $\phi_{n,m}$ with $m,n=P,T$ denotes the pion meson LCDAs deduced by the peudoscalar and tensor currents.
In the full amplitude $G$,
the gauge dependence proportional to $k_T$ from the two-parton Fock state cancels with
the gauge dependence associated with the three-parton Fock state,
then the gauge invariant LCDAs $\phi$ defined by the approximate path of Wilson link
deduces to the gauge invariance of the NLO (to all order) hard kernel $H(\zeta, b)$ \cite{FengZS,LiHU}.
The three-particle corrections to pion e.m. and $B \to \pi$ transition form factor are found at most a few percent \cite{ChenPN,ChenGV,ChengRUZ}.
The $k_T$ factorization and high power corrections are also surveyed in the $\rho$ meson involved processes
\cite{Zhang:2015mxa,ChengVNA,HuaKHO}.

It is easy to see from Eq.~(\ref{eq:NLO-KH}) that the NLO hard kernel is the difference between the NLO full amplitude and
the convolution of LO hard kernel with the NLO wave function,
retaining a scheme dependence to subtract the IR divergent contribution.
The analytical calculation is nontrivial because the partons are considered off shell by $k_T^2$
both in the quark diagrams from full QCD and in the effective diagrams of the wave function,
which introduces the double logarithms $\ln^2 k_T$ when the external parton momentum shrink in the IR region,
and the double logarithms $\ln^2 x$ associated with the on shell internal quark lines.
These two types of double logarithm are organized by $k_T$ and threshold resummations,
absorbed into the collinear/soft wave function and the hard jet function, repectively.
With setting both the factorization and the renormalization scales at the largest virtuality in the scattering,
the NLO QCD correction is found to be only a few percent to pion transition form factor \cite{NandiQX},
amount $20 \%$ to the pion e.m. form factor \cite{LiNN,ChengGBA},
and about $10 \%$ to the $B \to \pi$ transition form factor \cite{LiNK,ChengFWA}, the power accuracy is now up to two-parton twist three DAs of pion.

\subsection{TMD wave functions}

In the framework of collinear factorization theorem \cite{LangeFF},
a non-normalizable (unintegral) $B$ meson DA $\phi_+(k^+,\mu)$ is encountered in inclusive decays \cite{BraunWX},
here $k^+$ is the momentum carried by the light spectator quark.
In practice, this divergence ($ \sim 1/k^+$) does not break the collinear factorization at leading order
since only the first inverse moment $\lambda_B^{-1}(\mu) = \int dk^+ \phi_+(k^+,\mu)/k^+$ is relevant,
but the divergence emerges at high orders/powers with more moments interplaying,
which leads to an ambiguous renormalization of the decay constant $f_B$ \cite{LiJA}.

The TMD $B$ meson wave function defined under the HQET is
\beq
&~&\langle 0 \vert \, \bar{q}(y) \, W_y(n)^\dag \, I_{n,y,0}(n) \, W_0(n) \, \Gamma h(0) \, \vert \bar{B}(v) \rangle \, \non
&=&\frac{-if_Bm_B}{4} \mathrm{Tr} \Big[ \frac{1+\vsl}{2} \Big(2 \phi_+(v^+y^-, y_T^2)
+ \frac{\phi_+(v^+y^-, y_T^2) - \phi_-(v^+y^-, y_T^2)}{v^+y^-} \ysl \Big) \gamma_5 \Gamma \Big] \,,
\label{eq:B-WF}
\eeq
where $p_B = m_B v$,  the coordinate of field $\bar{q}$ is $y = (0, y^-, \mathbf{b})$,
$\phi_+$ and $\phi_-$ are the
are the leading and sub-leading twist DAs, respectively, corresponding to the
nonlocal matrix elements with $\bar{q}$ field moving along the light cone direction $n = n_? = (0, 1, \textbf{0}_T)$ with the momentum
fraction $x$ inside $B$ meson.
We remark here that the combination of these two leading twist DAs $\bar{\phi} \equiv \phi_+-\phi_-$
is numerically negligible \cite{KurimotoZJ,LuNY}.
The Wilson line operator  $W_y(n)$ is written in the path-ordered exponent
\begin{eqnarray}
W_y(n) = \mathcal{P} \, \mathrm{exp} \Big[ -i g_s \int_0^\infty d \lambda \, n \cdot A(y + \lambda n) \Big] \,.
\label{eq:Wilson-line}
\end{eqnarray}
There is an additional IR divergence, say, the light cone singularity, arose in the definition of TMD wave function
when the loop momentum is parallel to the Wilson line on the light cone \cite{CollinsFM}.
An artificial resolution with retaining the gauge-invariance is to rotate the Wilson line away from the light cone by an arbitrary direction,
and alleviate the factorization-scheme dependence by adhering it to a fixed off-shellness $n^2 \neq 0$.
This treatment introduces a scheme dependence (another scale) $\zeta_B^2 = 4 (n \cdot p_B)^2/n^2$ on the choice of $n$.
With introducing the gluon mass to regularize the soft divergence,
the DA $\phi_+(k^+, b, \mu)$ at NLO includes both
the UV logarithm $\ln(\mu b)$ which is summed to all order in an exponent $R$ by the standard RG equation,
and also the IR logarithms $\alpha_s \ln^2(\zeta_B b)$ and $\alpha_s \ln(\zeta_B b)$
which are resummed to the Sudakov factor $S$.
The evolution of DA $\phi_+$ is then given by
\begin{eqnarray}
\phi_+(k^+,b,\mu)  = S(k^+,b,\zeta_B)  \, R(b, \mu, \zeta_B) \, \phi_+(k^+,b,1/b) \,,
\end{eqnarray}
where the universal initial condition $\phi_+(k^+,b,1/b)$ absorbs all the soft logarithms in terms of gluon mass.
In the heavy quark limit, $b \to 0$ and the light cone divergences cancel each other.
The normalization of the DA $\phi_+(k^+, b, \mu)$ becomes realized in $k_T$ factorization
since the Sudakov evolution trends to identity in the limit $b \to 1/k^+$.
\begin{eqnarray}
\int_0^\infty dk^+ \,  \lim_{b \to 1/k^+} \, \phi_+(k^+, b , \mu)  = \int_0^\infty dk^+ \, R(1/k^+, \mu, \zeta_B) \, \phi_+(k^+, 1/k^+, \mu) \,.
\end{eqnarray}

Besides the light cone divergences associated with the $b$ parameter,   another type of large logarithms from the light-like Wilson line,
such as the rapidity singularity $\ln^2 (\zeta_B^2/m_B^2)$ and $\ln x \ln (\zeta^2_B/m_B^2)$,
emerges in the TMD definition when we do the NLO calculations to the exclusive $B$ meson decays.
As it is discussed for the $B \to \pi$ transition form factors \cite{LiNK,ChengFWA},
the arbitrary large value of $\zeta_B$ brings a factorization-scheme dependence,
which in principle should be resummed to all order by resolving the evolution equation of $B$ meson wave function
on the dimensionless scale $\zeta^2 \equiv \zeta_B^2/m_B^2$,
with considering simultaneously with the RG evolution in the factorization scale $\mu_f$.
The solution demonstrates that the resummation effect suppresses the shape of $B$ meson DA
$\phi_+(k^+,b,\mu)$ near the end point $k^+ \to 0$ \cite{LiMD}.
The third type of large logarithm related to the light cone singularity, in the mixed form $\ln x \ln(\zeta_\pi^2/k_T^2)$,
appears in the effective pion wave function at NLO \cite{LiNN,ChengGBA}, where $\zeta_\pi^2$ has the similar definition as $\zeta_B^2$.
To sum over all the mixed logarithm in pion wave function,
a joint resummation is constructed by the evolution equation on the rapidity parameter $\zeta_\pi^2/p^2_\pi$ \cite{LiXNA},
with combining the conventional threshold resummation and the $k_T$ resummation.
It is shown that the strong suppression at small $x$ and large $b$ is indeed happened,
the first scheme independent (of the light cone singularity) prediction of the pion form factors under $k_T$ factorization
reveals that the joint resummation improved pion wave function does not bring sizable corrections,
this is not accidental because the threshold and $k_T$ resummation have reshaped the end-point behaviours already.

Continuing on the study of light cone divergence in the $k_T$ factorization theorem,
the self-energy correction of the infinitely long dipolar Wilson links brings another pinch singularity \cite{Collins,JiWU},
which is suggested to be eliminated by introducing an auxiliary vacuum matrix element (soft function) and
supplementing a complicated soft subtraction in the TMD definition.
Recently, a new definition for TMD wave function is proposed
to solve the rapidity and self-energy divergences simultaneously in a much simpler formula \cite{LiXDA},
with comparing to the conventional Collins definition.
The key point is to rotate the Wilson links in the un-subtracted wave function away from the light cone (IR regulator),
and reverse the two-pieces of non-light-like Wilson links in different directions (non-dipolar Wilson links).
In the simplest case they are organized orthogonal to each other and the soft function is not needed again.
At NLO, both the conventional Collins definition and the non-dipolar definition yield to the same collinear logarithm $\ln^2 k_T$
as in the dipolar gauge links, the equivalence of these two definitions is illustrated by the same evolution kernel.

\subsection{Strong phases in the PQCD approach}

There are three sources of strong phase in the PQCD approach to analyze the nonleptonic two-body $B$ decays,
which can be read from the factorization formula in Eq.~(\ref{eq:PQCD}):
\begin{enumerate}
\item[(1)]
The first one comes from the Sudakov exponent,
which in principle relates to the angular distribution of scattering hadrons as well as the center of mass scattering angle.
In two-body hadronic $B$ meson decays this phase is negligible because of the color transparency that the final states are moving almost back to back,
but it may be important in the baryon decays with the angular distribution.

\item[(2)]
The second type of strong phase comes from the NLO correction to the spectator emission diagrams with the glauber gluons,
which supplements each of the final light mesons with a phase and
might modify the destructive interaction between different topological amplitudes to the constructive one at the tree level.

\item[(3)]
The last origin of strong phase is from the annihilation diagrams.
In the PQCD approach, the strong phase from the annihilation amplitudes  plays an important role to understand the large CPV for $B \to \pi^+\pi^-$
decays as reported  by B factories experiments \cite{Bfactory}.
This kind of strong phase in the PQCD approach, for example,  is very different from the one due to the Bander-Silverman-Soni mechanism  \cite{BSS1979}
in the QCDF  approach.
\end{enumerate}

In our classification,  the first two kinds of the strong phases associated with the wave functions
do reflect property of the soft and glauber gluon corrections at high orders,
while the strong phase from the  annihilation diagrams corresponds to the hard gluon exchanging in the timelike momentum regions.

The factorizable annihilation amplitudes is proportional to the timelike form factors of the interaction between the final two light mesons.
With the invariant mass of $m_B^2$ being far away from the resonances, the strong phase is expected to be around $\pi$ radian in this amplitude.
The principle value theorem of timelike propogator indicate a visibal imaginary part.
In the penguin-dominated $B$ decays, the imaginary part in the annihilation amplitude is at the same order ($\mathcal{O}(\alpha_s)$),
and also at the similar power as the emission amplitudes ($\mathcal{O}(\Lambda^4)$), then a large CP violation can be expected.
For the nonfactorizable annihilation amplitudes with the hard gluon emitting from $B$ meson to generate a sea quark pair in the final states,
the invariant mass decreases to $\mathcal{O}(m_B^2/4)$,
but the power counting in this amplitude is similar to the factorizable amplitude due to the Sudakov reshaping of the light mesons DAs.

In order to check the stability of PQCD prediction for strong phase in penguin dominated $B$ decay modes,
timelike pion e.m. form factor is studied with the help of analytic continuum transferred from spacelike to timelike regions \cite{HuCP,ChengQRA}.
The results show that the NLO radiative correction brings $20\%-30\%$ enhancement  to the magnitude in the region $q^2 > 10 \mathrm{GeV}^2$,
while decreases the strong phase by about $10$ degrees in the same region.
In fact, the strong phase in two-body $B$ decays arose from the annihilation amplitudes
with the timelike form factor deduced by the scalar current when the final states are identical particles \cite{NPB896-255}. 
It is worth of mentioning that in the large momentum transfer regions $Q^2 > 30 \mathrm{GeV}^2$,
the single-b convolution reproduces well for the result obtained in the conventional double-b convoluted formula for the timelike form
factor \cite{ChengKHI}, which verifies the fact that the internal hard gluon gives the dominate contribution to the strong phase and
the statement of hierarchy ansatz $x_i Q^2 \gg x_1x_2 Q^2 \sim k_{iT}^2$ in the PQCD approach.

\subsection{Hadronic $B_q$ decays at the LO PQCD approach}

Thanks to the great running of the $B$ factory and the LHC experiments in the new century,
several factorization approaches were established and developed to analyze the hard exclusive QCD processes.
During the past twenty years, the PQCD approach at the full LO are widely employed to predict
the two-body hadronic decays and the semileptonic decays of the $B_q$ mesons,
and help us to understand most of the experimental measurements for the branching ratios (BRs),
the CP violating asymmetries (CPV) and many other physical observables \cite{Bfactory,HFLAV19,pdg2020}.

At LO, the $B \to K \pi$  and $B \to \pi\pi$ decay modes \cite{KeumWI,LuEM} are firstly studied in the PQCD approach
with the inclusion of the large penguin enhancement effects and the non-negligible annihilation contributions.
Dozens of two-body charmless hadronic $B \to (PP, PV, VV)$ decays\footnote{We use the notations
$P, V, A, S, T$ to denote the pseudoscalar, vector, axial-verxtor, scalar and tensor light mesons, respectively.}
have been calculated \cite{LuHJ,LiTI,LiHG,Xiao-07a} and compared with the available measured results phenomenologically.
Some charmed B meson decays like $B \to ( D^{(\ast)}(P, V),  J/\psi K^\ast, D^{(\ast)}D^{(\ast)})$
are also studied following the experimental measurements \cite{KeumJS,LiXF}.
The charmless two-body decay modes of $B_s$ meson are also studied in Refs. \cite{70-034009,AliFF,ZouIWA}.
For the charmless pure annihilation decay $B_s^0 \to \pi^+\pi^-$,
for example, the LO PQCD predictions for its large branching ratio
${\cal B}(B_s^0 \to \pi^+\pi^-)\approx 5  \times 10^{-7}$ \cite{85-094003} are confirmed by the CDF and LHCb measurements \cite{HFLAV19,pdg2020}.

Considering the charmless two-body decays of double heavy meson $B_c$, only the annihilation mechanism plays in to the game.
About two hundred channels of this type decays are studied systematically in the PQCD with the predictions for their BRs,
the longitudinal polarization fractions and the angular observables \cite{Bc-2010a,Bc-2010b,Bc-2010c,Bc-14a}.
Furthermore, the decay mode with at least one charm meson or charmonium in the final states are also investigated,
i.e., the channels like  $B_c\to (\jpsi, \eta_c )(P,V) $ and $B_c \to X (D^{(*)}, S/A/T)$ \cite{Rui-2015,Rui-2016,Liu-2017,LiuKUO,tensor14},
where $X$ are the light or charmed mesons.

The three-body or quasi-two-body nonleptonic $B_q$ decays are studied in the resonance region,
corresponding to the intermediate parts of the edges in the Dalitz plot of final state phase space, by introducing two-meson wave functions.
For example, the leading power analysis of the branching ratio is investigated for $B^+ \to K^+\pi^+\pi^-$ decay \cite{ChenTH}
and the first PQCD prediction of the CP violation in the localized region of the phase space is carried out for the channels
$B^\pm \to \pi^+\pi^-\pi^\pm, \pi^+\pi^-K^\pm$ \cite{WangIRA},
which provide us an independent fashion to understand the measured sizable direct CP asymmetries
as reported by LHCb collaboration\cite{111-a,112-a}.
In the last five years, many quasi-two body hadronic decays  $B_q \to (\eta_c, D_{(s)}, K, \pi, \eta^{(')} ) (\pi\pi,K\pi, KK)$
are studied by including the major contributions from the resonant states $(f_0(X),\rho(X),K^*, \rm{etc.})$ for example in
Refs. \cite{Li-17a,Li-17b,Ma-17a,Li-20a,Wang-20a}.

 \subsection{Two-body hadronic $B_q$ decays and the NLO contributions}

\begin{figure}[tb]
\vspace{-5cm}
\centerline{\epsfxsize=19cm \epsffile{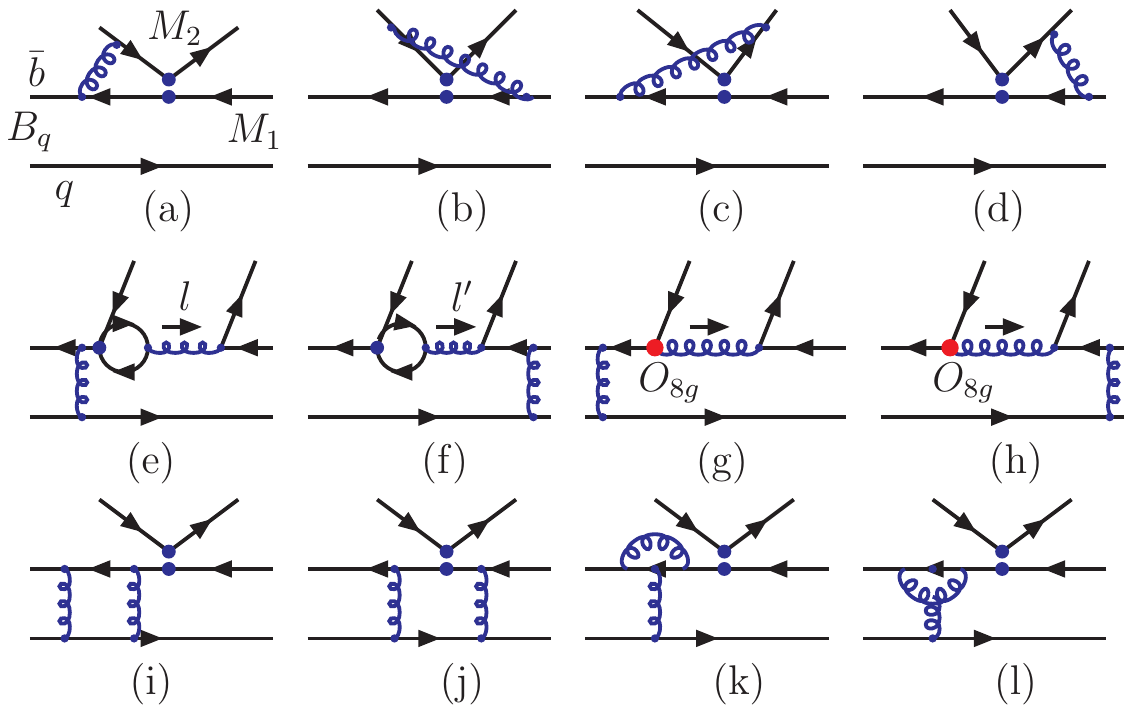}}
\vspace{-15cm}
\caption{The typical Feynman diagrams for the NLO contributions to $B_q\to M_1 M_2$ decays with
$q=(u,d,s)$ in the PQCD approach:  the vertex corrections  (a-d); the quark-loop contributions (e-f);
the chromo-magnetic penguin ${\cal O}_{8g}$ contributions (g-h);
the NLO twist-2 and twist-3 contributions to $B_{(s)} \to P$ transition form factors (i-l).}
\label{fig:fig2}
\end{figure}

\begin{figure}[tb]
\vspace{-5cm}
\centerline{\epsfxsize=19cm \epsffile{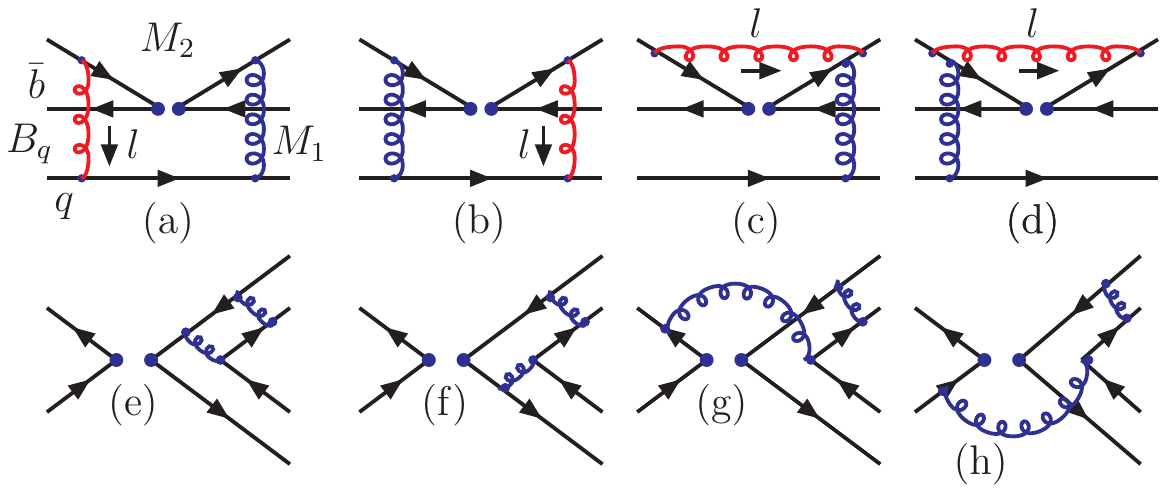}}
\vspace{-17cm} \caption{The typical Feynman diagrams for the NLO contributions to $B_q\to M_1 M_2$ decays
in the PQCD approach: the NLO contributions to spectators from the Glauber gluons (red curves)
which contains the Glauber divergences associated with $M_2$ meson (a-d); and the NLO contributions  to annihilation diagrams (e-h). }
\label{fig:fig3}
\end{figure}

In order to improve the reliability and accuracy of the PQCD approach,
ones have to take in to account the NLO contributions corresponding to the Feynman diagrams as illustrated in Fig.~\ref{fig:fig2} and \ref{fig:fig3},
where the sub-diagrams in each line of these two figures provide different kinds of NLO corrections
to the partner LO Feynman diagram depicited in Fig.~\ref{fig:fig1}.
At NLO level, one firstly should use:
(1) the NLO Wilson coefficients $C_i(\mu)$;
(2) the NLO renormalization group running matrix $U(m_1,m_2,\alpha)$ to run $C_i(\mu)$ from the high $\mu \sim M_W$ scale
to the low $\mu \lesssim m_b$ scale;
and (3) the strong coupling constant  $\alpha_s(\mu)$ at the two-loop level \cite{BuchallaVS}.

During past two decades, many authors have made great efforts to calculate the NLO contributions
to the two-body charmless hadronic $B_{(s)} \to M_1 M_2$ decays.
At present, most NLO contributions of $B/B_s \to P$  transitions in the PQCD approach have been evaluated.
\begin{enumerate}
\item[(1)]
The NLO vertex corrections (VC) to factorizable emission diagrams, as illustrated in Fig.~\ref{fig:fig2}(a-d),
were calculated under the collinear factorization hypothesis \cite{BenekeRY} without producing the end-point singularity.
and these results are then implemented directly in the PQCD approach \cite{LiKT} to deal with the charmless $B \to M_1 M_2$ decays.
This type corrections with a gluon exchanging between four quark effective operator can be absorbed into the Wilson coefficients $C_i(\mu)$,
regarding as a kind of corrections to the running of $C_i(\mu)$ from the high scale $m_W$ to the low  scale around $m_b$. 
Phenomenologically, it can produces a large enhancement to the decay amplitudes which is color suppressed at tree level,
and plays an important role in interpreting the so-called $\pi\pi, K\pi$ and $K\etapp$ puzzles \cite{LiKT}.

\item[(2)]
For the penguin-dominated decay modes, the quark-loop (QL) corrections as illustrated in Fig.\ref{fig:fig2}(e,f) provides sizable modifications to their total decay amplitudes \cite{LiKT}.
Because of the absence of end-point singularity, the transversal momentum can be dropped in the gluon invariant mass $l^2$ for simplicity
and the scale in factorization hypothesis can be expected at $l^2 \sim m_B^2/4$ in this case.
The QL corrections can be absorbed by introducing a new effective Hamiltonian $H^{QL}_{eff}$ compensated to Eq.\ref{eq:hamiltonian}.
Taking the branching ratios of $B \to K \pi $ decays for example,
a small enhancement ( $\sim 10 \%$ in size) is found by the up-loop and charm-loop corrections.

\item[(3)]
The third kind of NLO corrections is from the chromo-magnetic penguin (MP) operator ${\cal O}_{8g}$ shown in Fig.\ref{fig:fig2}(g-h),
whose contribution is firstly discussed in the PQCD approach for the tree-dominated $B \to \phi K$ mode \cite{MishimaWM},
and the further studies show that such correction induces a moderate changes to many considered $B$ meson decays \cite{WangMUA,Fan-13a,ZhangBSA}.
Once again, a new effective Hamiltonian $H_{eff}^{MP}$ is introduced to absorb this type contribution from MP correction.

\item[(4)]
Fig.~\ref{fig:fig2}(i-l) displays the NLO corrections to $B \to M$ transition form factors,
which are the basic physical quantities to construct the decay amplitudes of factorizable emission diagrams.
This part corrections have been calculated recently for the $B \to \pi$ transition with considering both
the two-particle twist-2 and twist-3 pion meson DAs \cite{LiNK,ChengFWA}.
The result shows that the NLO twist-3 correction is similar in size but has an opposite sign comparing to the NLO twist-2 correction. 
For the form factor $f_{+}^{B\to \pi}(0)$ associated with the Lorentz vectors summing over the momenta before and after decaying,
for example, the $24\%$ NLO twist-2 enhancement to the full LO prediction \cite{LiKT} is largely canceled by the $17\%$
NLO twist-3 reducing, leaving a small and stable $\sim 7\%$ correction.

\item[(5)]
Besides the NLO corrections to the form factors,
another interesting corrections to the hadron matrix elements  $\langle M_1M_2|O_i |B \rangle$
comes from the glauber gluon ($l \sim (\Lambda^2/m_B,\Lambda^2/m_B,\Lambda)$) \cite{LiHAA,LiuSRA,Liu-2016a}.
At the moment, this effect is concerned only in the nonfactorizable spectator diagrams,
as illustrated in Fig.~\ref{fig:fig3}(a-d) with the red curves denoting the glauber gluons.
The result show that only the glauber gluon from the light pion meson brings significant effect,
while the glauber gluon associated with the heavy $B$ meson is not important and can be ignored \cite{LiHAA}.
Saying more about the dynamics, the glauber effect formulates to an additional phase associated to the light meson $\pi$,
which effectively enhances the color-suppressed spectator tree amplitude and changes the interference mode between it with other tree amplitudes,
from destructive to instructive, and provides a possibility to understand the long-standing $\pi\pi$ puzzle. 

\item[(6)]
Fig.~\ref{fig:fig3}(e-f) presents the NLO corrections to factorizable annihilation amplitudes,
which is basically proportional to timelike meson form factors.
Concerning the $B \to \pi\pi$ process again, the timelike pion form factor deduced by the e.m current is largely cancelled
between Fig.~\ref{fig:fig3}(e) and (f) due to the identity of final two mesons (they are cancel to each other in the $\pi^0\pi^0$ case),
so this type amplitude receives contribution mainly from the timelike scalar pion form factor. 
In Ref.~\cite{NPB896-255}, we made the PQCD calculation of the NLO correction to the scalar pion form factor
and implemented it in the annihilation amplitude of $B \to \pi\pi$ decays.
We found that (a) the NLO part of the scalar pion form factor is very small in size but has has a large strong phase around $-55^\degree$,
which can produce a large CPV;
and (b) such NLO annihilation correction (${\cal O} (\alpha_s^2)$) produces a very small enhancement
to the branching ratios of $B \to (\pi^+\pi^-, \pi^0\pi^0)$ decays, less than $3\%$ in magnitude.
\end{enumerate}

The possible effects of the above NLO contributions have been examined in phenomenological studies for many
two-body charmless $B_{(s)} \to M_1M_2$ decays \cite{prd78-114001,cpc33-722,cpc38-033,npb931-79,npb935-17,npb946-114}
and the charmed $B_{(s)} \to (\jpsi, \eta_c ) (P, V) $ decays \cite{cpc34-937,npb953-114},
the agreements between the PQCD predictions with the measurements
for the BRs and other physical observables of those considered decay modes are improved effectively,
some so-called puzzles of $B \to \pi\pi, K\pi$ and $B \to K\etapp$ can be understood better with the NLO correction effects in the PQCD approach.

For the four penguin-dominated $B\to K \pi$ decays, for example, the measured large BRs can be understood successfully
in the PQCD with the inclusion of the partial NLO corrections (VC+QL+MP) known at 2005 \cite{LiKT},
but it is still very difficult to interpret the "$K\pi$" puzzle: why the measured values of the direct CPV $\acp^{\rm dir}(B^0\to K^\pm\pi^\mp)$
and $\acp^{\rm dir}(B^\pm \to K^\pm \pi^0)$ are so different ?
At the quark level, the $B^0 \to K^+\pi^-$ and $B^+ \to K^+ \pi^0 $ decays differ only by
the sub-leading color-suppressed tree ($"C"$) and the electroweak penguin ($"P_{EW}"$) operators,
so their direct CPV are generally expected to be similar,
one can see this straightly from the LO PQCD predictions in table \ref{tab:br1}.
By taking all the known NLO contributions into account, the PQCD predictions for both
$\acp^{\rm dir}(K^\pm\pi^\mp)$ and $\acp^{\rm dir}(K^\pm \pi^0)$ do become agree very well with the measurements.
As presented in table \ref{tab:br1}, the glauber gluon effects result in a small reduction to branching ratios ($\sim 10\% $)
while simultaneously flip the sign of $\acp^{\rm dir}(K^\pm \pi^0)$.
The underlying reason is that the $B \to K^\pm \pi^\mp$ decays are QCD-penguin dominated decay modes,
whose decay rates are not sensitive to $"C"$,
the $B^\pm \to K^\pm \pi^0$ decay, however, is sensitive to $"C"$ and can be modified evidently by the glauber gluon.

\begin{table}
\caption{The LO and NLO PQCD predictions for the BRs (in units of $10^{-6}$) and the direct CPV for $B\to K\pi $ decays \cite{LiKT,Liu-2016a},
where NLO (NLOG) denotes the predictions without (with) the glauber effects.
In the last column, the world averages values \cite{pdg2020} is presented.}
\vspace{1em}
\label{tab:br1}
\begin{tabular*}{16cm}{@{\extracolsep{\fill}}l|ll|ll|l} \hline\hline
{\rm Mode} & {\rm LO}\cite{LiKT} &  {\rm PQCD} \cite{LiKT}&  {\rm NLO}\cite{Liu-2016a} &  {\rm NLOG }\cite{Liu-2016a} & {\rm Data}\cite{pdg2020}  \\ \hline
 ${\cal B} (B^\pm \to K^\pm \pi^\mp)$&    $14.2  $ &$20.6^{+15.6}_{-8.3}$& $23.3^{+7.8}_{-5.7}$& $21.7^{+7.4}_{-5.3}$ & $19.6\pm 0.5  $\\
  ${\cal B} (B^\pm \to K^\pm \pi^0)$     &    $10.2  $ &$13.9^{+10.0}_{-5.6}$ & $15.3^{+5.2}_{-3.8}$& $14.0^{+4.7}_{-3.5}$ & $12.9\pm 0.5  $\\
 ${\cal B} (B^\pm \to K^0 \pi^\pm )$     &    $17.0  $ &$24.5^{+13.6}_{-8.1}$& $27.2^{+9.3}_{-6.7}$& $24.1^{+8.3}_{-6.0}$ & $23.7\pm 0.8  $\\
 ${\cal B} (B^0 \to K^0\pi^0)$                   &    $5.7     $ &$9.1^{+5.6}_{-3.3}$& $10.2^{+3.4}_{-2.5}$& $9.3^{+3.2}_{-2.3}$ & $9.9\pm 0.5  $\\ \hline
 $\acp^{\rm dir} (B^\pm \to K^\pm \pi^\mp)$&    $-0.12  $ & $-0.09^{+0.06}_{-0.08}$& $-0.076 \pm 0.017$ & $-0.081\pm 0.017  $ & $-0.083 \pm 0.004  $\\
  $\acp^{\rm dir} (B^\pm \to K^\pm \pi^0)$     &    $-0.08  $ & $-0.01^{+0.03}_{-0.05}$&$-0.008\pm 0.014$& $+0.021\pm 0.016  $ & $+0.037  \pm 0.021  $\\
 \hline \hline
\end{tabular*}
\end{table}

Although some great progresses have been reached on the NLO estimation
of the heavy-to-light transition form factors in the factorizable emission diagrams Fig.~\ref{fig:fig2}(i-l),
the timelike meson form factors in the factorizable annihilation diagrames Fig.~\ref{fig:fig3}(e-f)
and the glauber gluon effect in the non-factorizable spectator diagrams Fig.~\ref{fig:fig3}(a-d),
the NLO corrections to the non-factorizable annihilation diagrams in Fig.~\ref{fig:fig3}(g-h) are still absent.
Meantimes, the NLO calculations for the hard spectator diagrams have not been completed yet,
more time and efforts are clearly required. 

\subsection{Semileptonic $B_q$ decays in PQCD approach}

Besides the hadronic decays, the semileptonic B meson decays also play an important role in
the precision determination of the CKM matrix elements $\vert V_{ub} \vert$ and $\vert V_{cb} \vert$,
the understanding of the inner QCD structure of mesons,
as well as the search for the evidence of the new physics (NP) \cite{Bfactory,HFLAV19,pdg2020}. 
In the SM, the electroweak couplings of all leptons are equal except for the mass effects (i.e., phase space and helicity suppression),
which is one of the distinctive hypothesis of the SM and called the lepton flavor universality (LFU).
Any significant LFU violation will be a clear signal of the NP.

Since 2012, BaBar, Belle and LHCb collaborations have reported the so-called $R(D^{(\ast)})$ and $R(K^{(\ast)})$ anomalies
\cite{RD-12a,lhcb1308a,lhcb1308b,PRL122-191,RD-20a} that the measured ratios, defined as the ratios of the BRs
\beq
R(D^{(*)})&=&\frac{{\cal B}( \bar{B} \to D^{(*)} \tau^- \bar{\nu}_\tau )}{ {\cal B}( \bar{B} \to D^{(*)} l^- \bar{\nu}_l)},  \quad  {\rm for}  \quad l=(e,\mu) \,,
\label{eq:RD1}\\
R(K^{(*)})&=&\frac{{\cal B}( \bar{B} \to K^{(*)} \mu^+ \mu^-)}{ {\cal B}( \bar{B} \to K^{(*)} e^+ e^-)} \,,
\label{eq:RK1}
\eeq
disagree with the SM expectations \cite{PRD85-094,PRD88-074} by about $(2-4) \, \rm{\sigma}$.
The appearance of these anomalies, of course, invoked intensive studies for the various kinds of the semileptonic $B$ meson decays
in the SM and many NP models \cite{prl109-071,prl109-161,jhep2013-010,PRD95-115}.
For more details about the studies for these anomalies, one can see the recent review \cite{Li:2018lxi}, 
here we focus only on the systematic studies for various kinds of semileptonic decays of $B_q$ mesons in the PQCD approach
with the inclusion of some known NLO contributions. 

\begin{enumerate}
\item[(1)]
The $R(D^{(\ast)})$ anomaly is studied by employing the PQCD and the "PQCD+Lattice" approaches,
where the relevant $B \to D^{(\ast)}$ transition form factors in the low $q^2$ region of $0 \leq q^2 \leq m^2_\tau$ are calculated
in the PQCD framework and extrapolated to $q^2_{\rm{max}}$ by using the pole model parametrization \cite{csb59-125},
the available lattice QCD evaluations for the relevant form factors \cite{Lattice-19} at the high $q^2$ region
are taken as the input to improve the reliability of the extrapolation \cite{csb60-2009}, respectively. 
The "PQCD+Lattice" predictions for the $B \to D$ transition form factor
is compariable to the NLO light cone sum rule result at the large recoiled region \cite{Wang:2017jow},
and the predictions for the ratios $R(D^{(\ast)})$ agree with the world averages of the experimental measurements within $2 \, \rm{\sigma}$.
The $\bar{B}^0_s \to D_s^{(\ast)} l^- \bar{\nu}_l $ decays have the similar properties as the $B \to D^{(\ast)} l^-\bar{\nu}_l$ decays
because they have the same $b\to c l^- \bar{\nu}_l$ transitions at the quark level.
We therefore have defined and calculated the ratios $R(D_s^{(\ast)})$ and $R_{D_s}^{l, \tau}$ \cite{PRD89-014,cpc44-053},
suggesting the LHCb and Belle-II collaborations to measure these new observables.

\item[(2)]
The NLO corrections bring $10\%-20\%$ variations for the BRs of the semileptonic decays
$B_{(s)} \to (\pi, K, \etapp, G) ( l^+l^-, l\bar{\nu}_l, \nu \bar{\nu})$ in the framework of PQCD \cite{PRD86-114,BSL-13a},
resulting in a better agreement with the data.
To investigate the $R(K^{(\ast)})$ anomalies, the Bourrely-Caprini-Lellouch parametrization \cite{bcl09}
is employed to extrapolate the $B_s \to K^{(\ast)}$ form factors. 
The "PQCD+Lattice" prediction of ${\cal B}(\bar{B}_s \to K^* \mu^+ \mu^-)=(2.48^{+0.56}_{-0.50})\times 10^{-8}$ \cite{PRD102-013}
agrees well  with the LHCb measurement $(2.9\pm 1.1) \times 10^{-8}$ within errors\footnote{The
non-form-factor correction in the semileptonic $B$ decays are carried out in the QCD light cone sum rule,
where the long-distance effect generated by the four-quark operators with c-quarks
is calculated for $B \to K^{(\ast)} l^+l^-$ decays \cite{Khodjamirian:2010vf},
and the nonlocal contributions generated by the four-quark and penguin operators are discussed for
the $B \to K l^+l^-$ decays at large hadron recoiled region \cite{Khodjamirian:2012rm}.}.
Other  physical observables, like the ratios $(R_{K}^{e\mu }, R_{K^\ast}^{e\mu }, R_{K}^{\mu \tau},R_{K^\ast}^{\mu \tau})$,
the direct CPV ${\cal A}_{CP}$ of all considered decay modes, the longitudinal polarization asymmetry $P_L$,
the $K^\ast$ polarization fraction $F_L^{K^\ast}$ and the angular observables $P_{1,2,3}$ and $P^\prime_{4,5,6,8}$,  have been
studied too, and might be tested by future experimental measurements. 

\item[(3)]
The semileptonic charmness $B_c$ decays, i.e., $B_c \to D_{(s)}^{(\ast)}(l^+l^-, l \bar{\nu}_l, \nu \bar{\nu}) $ and
$B_c \to (\eta_c, \jpsi)  l \bar{\nu}_l $ are also studied \cite{PRD90-094,cpc44-023},
focusing on the observables like the BRs, the new defined ratios $R(D^{(\ast)}), (R_\eta,R_{\jpsi})$
and the longitudinal polarizations $(P_\tau(\eta_c),P_\tau(\jpsi))$. 
The PQCD predictions for the $B_c \to (D_{(s)}, D_{(s)}^\ast, \eta_c, \jpsi)$ transition form factors consist with those obtained by other methods.
The "PQCD+Lattice" prediction $R_{\jpsi} = 0.27\pm 0.01$ agrees with LHCb measured value $R_{\jpsi}=0.71\pm 0.24$
\cite{PRL120-121} within $2 \, \rm{\sigma}$. Other predictions could be tested by LHCb experiments in the near future.
\end{enumerate}

It is known that the tensions of ratios $R(D^{(*)})$ and $R(K^{(*)})$ still remain,
but become weaker than ever before when more events of the semileptonic
$B$ meson decays are recorded and analyzed by LHCb and Belle experiments.
For the case of $R(D)$ and $R(D^\ast)$,
although the world averages still have $\sim 3 \, \rm{\sigma}$ deviation from the SM, mainly due to the 2012 BaBar results,
the newest most precise Belle measurements with a semileptonic tagging method \cite{RD-20a}
\beq
R(D)&=&0.307\pm 0.037 (\mathrm{stat.}) \pm 0.016 (\mathrm{syst.}), \non
R(D^\ast)&=&0.283 \pm 0.018 (\mathrm{stat.}) \pm 0.014 (\mathrm{syst.}). \,
\eeq
do become consistent with SM predictions $R(D)=0.299 \pm 0.003$ and $R(D^*)=0.258 \pm 0.005$ \cite{HFLAV19}.
The Belle combined result agrees with the SM predictions within $1.6 \, \rm{\sigma}$  \cite{RD-20a}.
For the $R(K)$ and $R(K^\ast)$,
the tension between the SM predictions and the measurements before 2019 is $2.1-2.5 \, \rm{\sigma}$ in bins \cite{Mulder-20a}.
The newest LHCb measurement with "Run 1+ Run 2" data
is $R(K)=0.846^{+0.060}_{-0.054} (\mathrm{stat.})^{+0.014}_{-0.016} (\mathrm{syst.})$ \cite{PRL122-191},
which is compatible with the SM prediction $R(K) =1.00 \pm 0.01$ at $2.5 \, \rm{\sigma}$.
Future experimental measurements could tell us more.

\section{The key issues and the opportunities in the future}\label{Future}

To meet the high-precision frontier of the experimental measurements,
high power and high order corrections should be considered in the PQCD approach as much as possible.
More information of the hadron structures, such as the wave functions, can be explored by some nonperturbative approaches,
while the resummation effects of TMD wave functions for different types of hadrons should be addressed in $k_T$ factorization theorem.
In this section we would like collect the key problems we are facing to develop the PQCD approach with new vitalities,
and highlight the opportunities in the near future studies.

\subsection{The corrections from three-particle $B$ meson DAs}

The Sudakov exponentiation of double logarithms forces the virtual gluon to be far off-shell,
which means that the gluon propagates without radiating for a long distance $\sim \mathcal{O}(1/\Lambda)$.
The factorization theorem shows that the off-shell particle is more like to interact with the lowest Fock states of the initial and final state mesons,
while the corrections from the interactions with three-particle configurations are power suppressed by $\mathcal{O}(\Lambda/Q)$.

As an ideal channel to determine the $B$ meson wave functions,
the leptonic radiative decay $B \to \gamma l \bar{\nu}$ \cite{CharngFJ,Wang:2016qii}
is revisited in the PQCD approach with including the complete eight three-particle DAs of $B$ meson,
the subleading power hard kernel with two-particle $B$ meson DAs, as well as the hadronic structure of photon \cite{ShenABS}.
A half decrease correction is estimated respect to the leading contribution,
this variation is consistent with the light cone sum rules result \cite{Wang:2018wfj},
indicating the demand of a systematic investigation of power corrections and high order QCD radiation corrections simultaneously,
especially in the $B$ decays. 
In Refs. \cite{ChengRUZ,ShenVDC}, 
high power correction from pion meson DAs up to twist-four is replenished to pion e.m. and transition form factors,
showing that the contribution from the three-parton Fock state is at least one order of magnitude
smaller than that from the lowest Fock state, and confirming the convergence of twist expansion.
The 12 GeV upgrade program at Jefferson Lab will measure more data of $F_\pi$ in the intermediate energy regions,
providing a chance to extract the nonperturbative parameters of meson DAs,
i.e., the moments in Gegenbauer polynomials expansion, with the precise PQCD predictions.

Interplaying with lots of exclusive decay measurements, this type of power correction,
as well as the usually dropping terms proportional to $k_T$ in the hard kernel,
should be implemented to $B \to \pi, \rho, \cdots$ form factors.
In fact, the three particle $B$ meson DAs bring sizable correction to $B \to (P,V,S)$ transition form factor
from the LCSRs calculations \cite{ChengSMJ,LuCFC,Gubernari:2018wyi,Descotes-GenonBUD,Cheng:2019tgh}.
This power correction should be added to the PQCD framework
of non-leptonic charmless two-body $B$ decay to update the predictions, especially for the emission type amplitudes.
For the time-like form factor of the light mesons interaction emerged in the factorizable annihilation amplitudes,
we can make a comment that it is hardly affected by the power corrections
since the large invariant mass regions is already far away from the resonances
and the power corrections behaviours as $\mathcal{O}(1/m_B^2)$.

\subsection{NLO QCD radiative correction}

As the precise improvements parallel to the power corrections, the QCD radiative corrections should be completed process by process.
Here we suggest several exclusive processes which are important to help us to understand QCD and the factorization theorem.

\begin{itemize}
\item[(1)]
The leptonic radiative decay $B \to \gamma l \bar{\nu}$ \cite{CharngFJ,Wang:2016qii} with QCD radiation and the power corrections
is recommended to determine the $B$ meson DAs.
\item[(2)]
The $B \to \rho$ transition form factors at NLO is also needed, together with the $B \to \pi$ form factors \cite{LiNK,ChengFWA},
to give a combined constraint for the CKM matrix element $\vert V_{ub} \vert$.
A similar exploration is the NLO correction to $B \to D$ form factor to improve the determination of $\vert V_{cb} \vert$
\footnote{The most important CKM unitarity test is the unitarity triangle (UT), whereas a long standing tension is observed at the $\sim 3\sigma$ level
between the exclusive and inclusive determinations of the CKM matrix element $V_{ub}$ and $V_{cb}$.
To clarify this tension, an important direction of effort is to perform more accurate  calculations from the theoretical side,
both for the exclusive and inclusive processes.}.
\item[(3)]
The timelike form factors of light mesons with scalar and tensor currents
would contribute to the annihilation amplitudes at higher orders,
and provides another source of strong phases even though it is expected to be small in the $B$ decays.
\item[(4)]
The vertex correction in b quark decay (emission and annihilation) should be calculated independently in the $k_T$ factorization theorem,
rather than using directly the result obtained in the collinear factorization \cite{npb675-333}.
In this way at least double sub-diagrams would be involved.
\item[(5)]
Some other still missing NLO corrections in the spectator and annihilation diagrams,
like the glauber gluon in the non-factorizable annihilation amplitudes, 
which may influence significantly for the transition amplitudes of nonleptonic two-body $B$ meson decays
which are color suppressed at tree level.
\end{itemize}

We also would like to mark some important inclusive $B$ decay processes.
In the PQCD approach the decay rate of inclusive $B$ decay, taking $B \to X_u l \bar{\nu}$ for example \cite{LiGI},
is written as a convolution of the hard amplitude, the $B$ meson wave function and the $u$-quark jet function,
in which the double logarithm emerged at end-point in the jet function is resummed into a Sudakov factor again.
Here the only theoretical input is the wave function of $B$ meson, whose first and second moments in the HQET definition,
Eq.(\ref{eq:B-WF}), can be determined from the photon energy spectrum of the nonleptonic radiative decays.
Considering the resummation effect in the penguin induced inclusive decay $B \to X_s \gamma$ \cite{LiKN2},
and also in the exclusive decays $B \to K^\ast \gamma, D^{(\ast)}\pi$ \cite{Melic},
a sharp peak at the low momentum fraction $x$ (equivalent to the moment $\lambda_B = 0.462 \, \mathrm{GeV}$)
and a moderate shape ($\lambda_{D^{(\ast)}} = 0.8 \,\mathrm{GeV}$) are obtained for $B$ and $D^{(\ast)}$ meson wave functions,
respectively, by fitting to the data.
The BRs of flavor-changing-neutral-current (FCNC) process deduced by $b \to d \gamma$ transition
can be expected to be suppressed by $\mathcal{O}(10^{-2})$ with respect to that of $b \to s \gamma$ transition,
whereas the PQCD predictions for $B \to \rho^0 \gamma, \omega \gamma$ \cite{LuYZ} are
at least two times larger than the measured ones \cite{pdg-2018}, leaving a big room for improvement.
Meanwhile, the PQCD predictions for the annihilation type radiative decays $B \to \phi \gamma, J/\psi \gamma$ \cite{LiXE}
are waiting to be tested by the experiments.
One can see that  (a) all these PQCD calculations of the inclusive $B$ decays
are still at leading order and need to be promoted to NLO accuracy in the $k_T$ factorization theorem;
(b) the $b \to s g$ transition via $s/g\to \gamma$ fragmentation  functions should be estimated,    which may be important for the $B \to X_s \gamma$ decay;
and (c) the formalism to organize the large corrections in three-particle ($qg\bar{q}$) contribution in a hard scattering is also interesting
to study the photon processes.

\subsection{TMD wave functions of $B$, $B_c$ and $\Lambda_b$}

At LO, both the leading and sub-leading twist $B$ meson DA, say, $\phi_+$ and $\phi_-$, contribute to $B \to \pi$ form factors,
by convoluting with leading and sub-leading twist pion meson DAs, respectively. 
The numerical calculation shows that their contributions are at the same order of magnitude in the moderate momentum transfer regions,
due to the chiral enhancement effect of pion meson DAs at sub-leading twist \cite{ChengFWA}.
So it is urgent to do the Sudakov resummation for the $k_T$ logarithms related to the DA $\phi_-(k^+, b , \mu)$,
as what have been done for the DA $\phi_+(k^+, b , \mu)$,
even though $\phi_-(k^+, b , \mu)$ does not appear in the $k_T$ factorization asymptotic formula \cite{LiJA}.
The improved shape of $\phi_-$ may help us to recheck, in an independent way,
the surprisingly large NLO correction to $B \to \pi$ form factor proportional to the DA $\phi_+$ \cite{LiNK},
and would help us to know well for the decay constant $f_B$ defined in the $k_T$ factorization for the normalizable $B$ meson wave function.

As the unique bound state composed of two different heavy flavor quarks,
$B_c$ meson can only decay via weak interactions.
The pioneering study of $B_c$ wave function is performed under the nonrelativistic potential model
by considering the relative motion between the two heavy quarks \cite{ChangPT},
and the hadron matrix elements in $B_c$ decays is also suggested to be calculated in the instantaneous nonrelativistic approximation \cite{ChangGN}.
To examine the applicability of the HQET in describing this double heavy system,
the pure annihilation decays $B_c \to (PP,PV,VV)$ and some of the $b$ quark decay modes $B_c \to D^{(\ast)}_{(s)} (P,V,T)$
have been studied in the PQCD approach \cite{Bc-14a}.
These predictions are obtained by implementing the conventional PQCD framework constructed for $B$ meson decays,
supplementing with the effect of charm quark mass in the hard kernels.
In Ref. \cite{LiuKUO} an improved $B_c$ DAs is speculated through the golden channel $B_c \to J/\psi \pi$
by substituting both the lower bounds of factorization and renormalization scales
from $1/b$ to $m_c$ in the Sudakov exponent of the energetic charm quark and the RG evolution of $B_c$ meson DAs, respectively.
In order to have a rigorous formalism of $B_c$ decays in the PQCD appraoch,
the $m_c$ effect should also be considered in the $k_T$ resummation for
$B_c$ meson wave function \cite{Liu:2020upy} as well as the $m_b$ effect which has already be studied.
A complete one-loop computation and an exact NLO resummation associated with an energetic massive quark is indispensable in the future work.

The PQCD approach is also applied to the semileptonic baryon decays $\Lambda_b \to p l \bar{\nu}$ \cite{ShihPB},
the reliable result is got only at the high end of the proton energy after taking into account the Sudakov resummation for $\Lambda_b$,
and the PQCD prediction of the $\Lambda_b \to p$ form factors \cite{HeUD,LuCM} are at least an order of magnitude smaller than the results
obtained by using the nonperturbative approaches \cite{WangSM,Khodjamirian:2011jp,Wang:2009hra,Wang:2015ndk},
indicating that the factorizable contribution may be negligibly small.
A concession to understand this unexpected result is to postulate the soft dynamics of $\Lambda_b \to p$ form factors
by making compromise that this form factor can not be calculated with the hard scattering theory,
and a hybrid PQCD approach was proposed to calculate the charmless hadronic decays of $\Lambda_b \to p \pi^-, p K^-$
by taking the form factors as nonperturbative input \cite{LuCM}.
In the hybrid PQCD approach, the nonfactorizable amplitudes are still dominated even though the factorizable amplitudes are no longer negligible,
implying a query that does the nonfactorizable amplitudes of the considered $\Lambda_b$ decays could be computed correctly in this hybrid PQCD approach,
in other word, does the formulism in $k_T$ factorization is still hold for $b$ baryon decays ?
We later realize that this deficiency may be caused by two reasons:
one is the DAs of proton employed in the PQCD calculation are motivated by the quark model, which is not constrainted by QCD,
another one is the simple implantation of the Sudakov exponent for $B$ meson in the $\Lambda_b$ baryon modes,
which are different in principle.
So the revisiting on heavy-to-light baryons transition with the QCD-based baryon LCDAs and the
perturbative resummation of $\Lambda_b$ baryon wave function is an important subject in the future study.

\section{Summary}\label{Summary}

Starting with the resummation technique and its application in exclusive QCD processes,
we make a brief review of the development of the PQCD approach and its recent progresses towards to the NLO accuracy
in the systematic studies for various kinds of $B/B_s/B_c$ meson decays.
We take the pion e.m. form factor as example to show the importance of transversal momentum with eliminating the end-point singularity
in elastic scattering processes by $k_T$ resummation.
The ideas embodying both the resummation effect and the hard scattering mechanism is then implemented to $B$ decays to explain the data,
and the PQCD approach is well-formulated in the process of data-promotion.
We concentrate on the recent progresses of the PQCD approach towards to NLO accuracy.
The factorization proofs and NLO calculations of several form factors are demonstrated under the $k_T$ factorization theorem,
the TMD definition of $B$ meson wave function and the light-cone singularities are emphasized,
and the NLO effects in nonleptonic and semileptonic $B$ decays are discussed.
Moreover, we show the possible strong phases would be generated in the PQCD approach,
where the annihilation amplitudes proportional to timelike form factors of light mesons are discussed in detail.
We also list some key problems which needed to be resolved to promote the development of the PQCD approach.
Higher order and higher power QCD corrections are indispensable,
the resummation effects in doubly heavy meson and baryon are also pressing.

\section*{Acknowledgments}

We are grateful to Yue-Long Shen, Ying-Ying Fan, Jun Hua, Xin Liu, Wen-Fei Wang and Ya-Lan Zhang
for the long-term collaborations and fruitful discussions on the NLO progresses of the PQCD factorization approach,
and especially to Hsiang-nan Li, Cai-dian L\"u and Yu-ming Wang for the careful reading and helpful comments on the manuscript.
This work is supported by the National Science Foundation of China (NSFC) under Grant No. 11805060, 11975112 and 11775117.

\end{document}